\documentclass{aa}
\usepackage{txfonts}
\usepackage{graphicx}
\usepackage[authoryear,longnamesfirst]{natbib}
\bibpunct{(}{)}{;}{a}{}{,}
\usepackage{longtable}
\begin{document}
%


\title{The young stellar population in the Serpens Cloud Core: \\ 
       An ISOCAM survey\thanks{Based on observations with ISO, an ESA 
                               project with instruments funded by ESA Member 
                               States (especially the PI countries: France, 
                               Germany, the Netherlands and the United 
                               Kingdom) and with participation of ISAS and 
                               NASA.}
                   \thanks{Tables~\ref{tbl-2} and \ref{tbl-3} are only
                           available in electronic form at the CDS via
                           anonymous ftp to cdsarc.u-strasbg.fr (130.79.128.5)
                           or via http://cdsweb.u-strasbg.fr/cgi-bin/qcat?J/A+A/}
       }

\author{A.A. Kaas\inst{1,2} \and 
        G. Olofsson\inst{2} \and  
        S. Bontemps\inst{3,2} \and 
        P. Andr\'e\inst{4} \and    
        L. Nordh\inst{2} \and  
        M. Huldtgren\inst{2} \and 
        T. Prusti\inst{5} \and  
        P. Persi\inst{6} \and 
        A.J. Delgado\inst{7} \and
        F. Motte\inst{4} \and
        A. Abergel\inst{8} \and 
        F. Boulanger\inst{8} \and 
        M. Burgdorf\inst{14} \and 
        M.M. Casali\inst{9} \and  
        C.J. Cesarsky\inst{4} \and 
        J. Davies\inst{10} \and 
        E. Falgarone\inst{11} \and 
        T. Montmerle\inst{12} \and 
        M. Perault\inst{11} \and 
        J.L. Puget\inst{8} \and
        F. Sibille\inst{13}}

\institute{Nordic Optical Telescope, Apdo 474, 38700 Santa Cruz de La Palma, Spain
            \and Stockholm Observatory, Roslagstullsbacken 21, 10691 Stockholm, 
                 Sweden 
            \and Observatoire de Bordeaux, BP89, FR-33270 Floirac, France
            \and Service d'Astrophysique, CEA Saclay, 91191
                 Gif-sur-Yvette,France
            \and Research and Scientific Support Department of ESA, Postbus 229,
                 2200 AG Noordwijk, The Netherlands
            \and Instituto Astrofisica Spaziale e Fisica Cosmica, CNR, Rome, Italy
            \and Instituto de Astrof\'isica de Andalucia, Granada, Spain
            \and IAS, Universit\'e Paris XI, 91405 Orsay, France
            \and Royal Observatory, Blackford Hill, Edinburgh EH9 3HJ, UK
            \and Joint Astronomy Centre, 660 N. A'Ohoku Place, University
                 Park, Hilo, HI 96720, USA
            \and ENS Radioastronomie, 24 Rue Lhomond, 75231 Paris, France
            \and Laboratoire d'Astrophysique de Grenoble, 38041 Grenoble Cedex, 
                 France
            \and Observatoire de Lyon, 69230 Saint Genis Laval, France
            \and SIRTF Science Center, California Institute of Technology,
                 220-6, Pasadena, CA 91125
           }

\offprints{A.A. Kaas}
\mail{akaas@not.iac.es}

\date{Received date; Accepted date}

\titlerunning{ISOCAM survey in Serpens}
\authorrunning{Kaas et al.}

\abstract{We present results from an ISOCAM survey in the two broad band 
filters LW2 (5-8.5 $\mu$m) and LW3 (12-18 $\mu$m) of a 0.13 square degree
coverage of the Serpens Main Cloud Core. A total of 392 sources were 
detected in the 6.7 $\mu$m band and 139 in the 14.3 $\mu$m band to a 
limiting sensitivity of $\sim$ 2 mJy. We identified 58 Young Stellar
Objects (YSOs) with mid-IR excess from the single colour index $[14.3/6.7]$, 
and 8 additional YSOs from the $H-K/K-m_{6.7}$ diagram. 
Only 32 of these 66 sources were previously known to be YSO candidates. 
Only about 50\% of the mid-IR excess sources show excesses in the near-IR 
$J-H/H-K$ diagram. 
In the 48 square arc minute field covering the central Cloud Core the
Class\,I/Class\,II number ratio is 19/18, i.e. about 10 times larger than 
in other young embedded clusters such as $\rho$ Ophiuchi or Chamaeleon. 
The mid-IR fluxes of the Class\,I and flat-spectrum sources are found to 
be on the average larger than those of Class\,II sources. Stellar 
luminosities are estimated for the Class\,II sample, and its luminosity 
function is compatible with a coeval population of about 2 Myr which 
follows a three segment power-law IMF. For this age about 20\% of the 
Class\,IIs are found to be young brown dwarf candidates. The YSOs are in 
general strongly clustered, the Class\,I sources more than the Class\,II 
sources, and there is an indication of sub-clustering. The sub-clustering 
of the protostar candidates has a spatial scale of 0.12 pc. These 
sub-clusters are found along the NW-SE oriented ridge and in very good 
agreement with the location of dense cores traced by millimeter data. The 
smallest clustering scale for the Class\,II sources is about 0.25 pc, 
similar to what was found for $\rho$ Ophiuchi. Our data show evidence 
that star formation in Serpens has proceeded in several phases, and that 
a ``microburst'' of star formation has taken place very recently, probably 
within the last 10$^5$ yrs.

}

\maketitle


\section{Introduction}
\label{intro}

The youngest stellar clusters are found deeply embedded in the molecular 
clouds from which they form. There are several reasons why very young 
clusters are particularly interesting for statistical studies such as mass 
functions and spatial distributions. 
Because mass segregation and loss of low mass members due to dynamical 
evolution has not had time to develop significantly for ages $\la 10^8$ 
yrs \citep{sca98}, the stellar IMF can in principle be found for the 
complete sample, at least for sufficiently rich clusters.
For ages $\la 10^5$ yrs the spatial distribution should 
in gross reflect the distribution at birth, which gives important input 
to the studies of cloud fragmentation and cluster formation. Only in the
youngest regions of low mass star formation do we find the co-existence
of newly born stars and pre-stellar clumps, which allows one to compare
the mass functions of the different evolutionary stages. Low mass stars
are more luminous when they are young, being either in their 
protostellar phase or contracting down the Hayashi track, which 
permits probing lower limiting masses.  
Severe cloud extinction, however, requires sensitive IR mapping at high 
spatial resolution to sample the stellar population of embedded 
clusters.

ISOCAM, the camera aboard the ISO satellite \citep{kes96}, provided 
sensitivity and relatively high spatial resolution in the mid-IR \citep{ces96}.
The two broad band filters LW2 (5-8.5 $\mu$m) and LW3 (12-18 $\mu$m), 
designed to avoid the silicate features at 10 and 20 $\mu$m, were selected 
to sample the mid-IR Spectral Energy Distribution (SED) of Young Stellar 
Objects (YSOs) in different evolutionary phases. According to the current 
empirical picture for the early evolution of low mass stars
\citep{ada87,lad87,and93,and94},
newborn YSOs can be observationally classified into 4 main evolutionary 
classes. Class\,0 objects are in the deeply embedded main accretion phase 
($\ga$ 10$^4$ yrs), and have measured circumstellar envelope masses 
larger than their estimated central stellar masses, with overall SEDs 
resembling cold blackbodies and peaking in the far-IR. Class\,I sources 
($\sim 10^5$ yrs) are observationally characterised by a broad SED with a 
rising spectral index\footnote{The spectral index is defined 
as $\alpha_{\rm IR} = d \log (\lambda F_{\lambda})/(d \log \lambda)$ and 
is usually calculated between 2.2 $\mu$m and 10 or 25 $\mu$m.} towards 
longer wavelengths ($\alpha_{\rm IR} > 0$) in the mid-IR.  The Class\,II 
sources spend some 10$^6$ yrs in a phase where most of the circumstellar
matter is distributed in an optically thick disk, displaying broad SEDs 
with $-1.6 < \alpha_{\rm IR} < 0$. At $\alpha_{\rm IR} \approx -1.6$ 
the disk turns optically thin, and the sources evolve into the 
($\sim 10^7$ yrs) Class\,III stage where the mid-IR imprints of a disk 
eventually disappear. A normal stellar photosphere has $\alpha_{\rm IR} 
= -3$. Thus, while Class\,0 objects are not favourably traced by mid-IR
photometry, they are expected to be rare. At the other extreme, Class\,III 
sources cannot generally be distinguished using mid-IR photometry 
since most of them have SEDs similar to normal stellar photospheres. 
But mid-IR photometry from two broad bands, as obtained in this study with 
ISOCAM, is highly efficient when it comes to detection and classification 
of Class\,I and Class\,II sources. Thus, considering the fact that these 
latter objects constitute the major fraction of the youngest YSOs, the 
ISOCAM surveys provide a better defined sample for statistical studies 
than e.g. near-IR surveys for regions with very recent star formation 
\citep[see][]{pru99}.

This paper presents the results from an ISOCAM survey of $\sim $ 0.13 square 
degrees around the Serpens Cloud Core in two broad bands centred at 6.7 and 
14.3 $\mu$m. This cloud, located at $b^{\rm II} = 5^{\circ}$ and $l^{\rm II} 
= 32^{\circ}$ at a distance of $259 \pm 37$ pc \citep{str96,fes98}, comprises 
a deeply embedded, very young cluster with large and spatially inhomogeneous 
cloud extinction, exceeding 50 magnitudes of visual extinction. Only a few 
sources are detected in the optical \citep{har85,gom88,gio98}. Serpens contains 
one of the richest known collection of Class\,0 objects 
\citep{cas93,hur96,wol98,dav99}, an indication that this cluster is young and 
active. On-going star formation is also evident from the presence of several 
molecular outflows \citep{bal83,whi95,hua97,her97,dav99}, pre-stellar 
condensations seen as sub-mm sources \citep{cas93,mcm94,tes98,wil00}, a 
far-IR source (FIRS1) possibly associated with a non-thermal triple radio 
continuum source \citep{rod89,eir89,cur93}, a FUor-like object \citep{hod96}, 
and jets and knots in the 2.1 $\mu$m H$_2$ line \citep{eir97,her97}. 
Investigations of the stellar content have been made with near-IR surveys 
\citep{str76,chu86,eir92,sog97,gio98,kaa99a}, identifying YSOs using 
different criteria, such as e.g. near-IR excesses, association with 
nebulosities, and variability. 

In this paper we identify new cluster members, characterize the YSOs into 
Class\,I, flat-spectrum, and Class\,II sources, estimate a stellar luminosity 
function for the Class\,II sample and search for a compatible IMF and age, and 
finally describe the spatial distribution of both the protostars and the 
pre-main sequence population in this cluster.
 
{\small
\begin{table*}
\caption[]{Observational parameters, detection statistics and photometric
           results for each of the 6 ISOCAM rasters CE, CW, CS, D1, D2, D3. 
           See Fig.~\ref{figmap} and text. \label{tbl-1}}

\[
\begin{tabular}{lcccrrcrrrrrr}
\hline
\noalign{\smallskip}

   & $\alpha$(2000) & $\delta$(2000) & Size$^1$ & T$_{\rm int}$ & pfov & 
   $<$n$_{\rm ro}>$ & N$_{\rm det}$ & N$_{\rm 6.7}$ & N$_{\rm 14.3}$ & 
   N$_{\rm both}$ & 1$\sigma_{6.7}$ & 1$\sigma_{14.3}$  \\
   &   &  & (\arcmin $\times$ \arcmin) & (sec) & (\arcsec) &  &  & & & & 
   (mJy) & (mJy) \\
(1) & (2) & (3) & (4) & (5) & (6) & (7) & (8) & (9) & (10) & (11) & (12) & 
(13) \\
\hline
CE    & 18 29 48.3 & 01 16 04.5 &13$\times$13  & 0.28& 3/6 &4$\times$13& 133 & 113 & 56 & 51 & 2 &3 \\
CW    & 18 29 06.3 & 01 16 01.4 &13$\times$13  & 2.1 & 6   &7          & 152 & 139 & 43 & 38 &0.8&2  \\
CS    & 18 29 52.5 & 01 02 37.9 &13$\times$16.5& 0.28& 6   &4$\times$13& 165 & 152 & 41 & 34 &1.2&4  \\
D1    & 18 29 48.7 & 01 15 20.5 &1.8$\times$4.6& 2.1 & 3   &44         &  20 &  15 &  9 &  9 & 1 &2  \\
D2    & 18 29 52.3 & 01 15 20.8 &1.8$\times$4.6& 2.1 & 3   &44         &  24 &  19 & 11 & 10 & 1 &4.5  \\
D3    & 18 29 57.8 & 01 12 55.4 &4.6$\times$1.8& 2.1 & 3   &44         &  30 &  21 & 15 & 13 & 1 &3 \\
Tot$^2$ &            &            &              &     &     &           & 421 & 392 &140 &124 &   &  \\

\noalign{\smallskip}
\hline
\end{tabular}
\]
$^1$ Approximate size since each raster has tagged edges. \\
$^2$ The total number is corrected for multiply observed sources in 
overlapped regions. \\

\end{table*}
}
%
\section{Observations and reductions}
\label{obs}

\subsection{ISOCAM}

This work presents data from the two ISOCAM star formation surveys 
LNORDH.SURVEY\_1 and GOLOFSSO.D\_SURMC. These were surveys within the 
ISO central programme mapping selected parts of the major nearby star 
formation regions in the two broad band filters LW2 (5-8.5 $\mu$m) and 
LW3 (12-18 $\mu$m). Other results on young stellar populations based
on these surveys comprise the Chamaeleon~I, II and III regions 
\citep{nor96,olo98,per00}, the $\rho$ Ophiuchi star formation region 
\citep{bon01}, the R Corona Australis region \citep{olo99}, and the
L1551 Taurus region \citep{gaa04}, all observations performed basically
in the same way. Data reduction methods were generally the same for
all regions, using a combination of the CIA\footnote{A joint 
development by the ESA Astrophysics Division and the ISOCAM Consortium 
led by the ISOCAM PI, C. Cesarsky, Direction des Sciences de la 
Mati\`{e}re, C.E.A., France} (CAM Interactive Analysis) package and own
dedicated software. For overviews of survey results, see \citet{nor98}, 
\citet{kaa00}, and \citet{olo00}. 

In the Serpens Core region about 0.13 square degrees (deg$2$) were covered 
at 6.7 and 14.3 $\mu$m in 3 separate, but overlapping main rasters named 
CE, CW and CS (see Fig.~\ref{figmap}). 
In addition, 3 smaller fields within CE (named D1, D2, and D3) were observed 
at a higher sensitivity. CE covers the well known Serpens Cloud Core, CW is 
a reference region to the west of the Core which is substantially opaque in 
the optical but without appreciable 60 $\mu$m IRAS emission \citep{zha88a}, 
and CS is a region directly to the south of the Cloud Core which has a peak 
in the 60 $\mu$m flux. 
The rasters were always made along the right ascension, with about half 
a frame (90\arcsec) overlap in $\alpha$ and 24$\arcsec$ overlap in $\delta$. 
For the larger rasters the pixel field of view (PFOV) was set to the nominal 
survey value of 6$\arcsec$ except for the CE region in LW2, where a PFOV of 
3$\arcsec$ was selected because of the risk of otherwise saturating the detector. 
Also in order to avoid saturation, the intrinsic integration time was set to 
$T_{\rm int} = 0.28$ s except for field CW where the nominal survey value of 
$T_{\rm int} = 2.1$ s was used. Each position in the sky was observed during 
about 15 s. For the deeper imaging within CE the overlap was 72$\arcsec$ in 
$\alpha$ and $\delta$, and each position was observed during about 92 s. 
The individual integration time was $T_{\rm int} = 2.1$ s and a PFOV of 
3$\arcsec$ was used for better spatial resolution. See Table~\ref{tbl-1} for
an overview.

\subsubsection{Image reduction}

Each raster consists of a cube of frames which is reduced individually,
and in total 9112 individual frames were analysed.
The dark current was subtracted using the CAL-G dark from the ISOCAM 
calibration library and, if necessary, further improved by a second 
order dark correction using a FFT thresholding method \citep{sta96a}. 
Cosmic ray hits were detected and masked by the multiresolution median 
transform (MMT) method \citep{sta96b}. The transients in the time 
history of each pixel due to the slow response of the LW detector were 
treated with the IAS inversion method v.1.0 \citep{abe96a,abe96b}.
Flat field images were constructed from the observations themselves.
See \citet{sta99} for a general description of ISOCAM data processing.

\subsubsection{Point source detection and photometry}
\label{obs-phot}

Bright point sources, even well below the saturation level, produce strong
memory effects which are not entirely taken out by the transient correction.
Due to the large PFOV which undersamples the point spread function (PSF), 
also remnants of cosmic ray glitches may be mistaken for faint sources.
By looking at the time history of the candidate source fluxes, however,
it is easy to distinguish real sources from memory effects, glitches or
noise. This is done for each individual sky coverage, the redundancy being
2-6. Both the source detection and the photometry was made with the interactive 
software developed by our team. See also previous papers on this survey 
\citep{nor96,olo99,per00,bon01}. 

The fluxes of each verified source were measured by aperture photometry, 
with nominal aperture radii 1.5 and 3 pixels, for the 6$\arcsec$ and 
3$\arcsec$ PFOV, respectively, and aperture correction applied using the 
empirical PSF. For each redundant observation (2-6 overlaps) the flux 
is the median flux from the $N_{\rm ro}$ readouts per sky position. The 
sky level was estimated from the median image to reduce the noise. 
Uncertainties are estimated both from the standard deviation around this 
median (i.e. the temporal noise and also a measure of the efficiency of 
the transient correction method) and from the $\sigma$ of the sky 
background (i.e. the spatial noise). The quoted flux uncertainties are 
these two contributions added in quadrature. 
The photometric scatter between the overlaps is estimated and deviating measurements 
are discarded. If a source is affected by the dead column (no 24 was 
disconnected) or is hit by a serious glitch or memory effect in one of 
the redundant overlaps, this one is skipped, and the remaining ones are used 
to estimate the flux. If the redundancy is not sufficient (i.e. along raster 
edges), then the fluxes are flagged if affected by the dead 
column, detector edges, glitches, memory effects, close neighbours etc. 

Source positions were calculated from that of the redundant images where 
the source is closest to the centre, in order to diminish the effects of 
field distortion. The source centre is taken to be the peak pixel, and 
the default ISO pointing was used as a first source position value. 
The ISO position was then compared to near-IR positions of known sources 
in the literature \citep{eir92,sog97,gio98}, and ISO positions of bright 
optical sources to the digital sky survey. Bulk offsets found for each 
raster registration were then corrected. The maximum bulk offset found 
was 5$\arcsec$, and we estimated an average uncertainty of $\pm 3\arcsec$
in RA and DEC. When 2MASS became available we checked the positions of
61 sources in Table~\ref{tbl-2} which have 2MASS counterparts and are not 
multiples unresolved by ISOCAM. The median deviation between ISOCAM and 2MASS 
positions is $2.2\arcsec$. One star deviates by as much as $9.2\arcsec$ 
(ISO-356), and 5 more sources by more than $4.5\arcsec$ (these are ISO-29, 
207, 272, 357 and 367).

\subsubsection{Photometric Calibration}

The two broad band filters LW2 and LW3 have defined reference wavelengths
at 6.7 and 14.3 $\mu$m, respectively. The fluxes in ADU/s are converted to 
mJy through the relations 2.32 and 1.96 ADU/gain/s/mJy for the bands LW2 and 
LW3, respectively (from in-orbit latest calibration by \citet{blo00}). These 
flux conversions are strictly valid only for sources with F$_{\nu} \propto 
\nu^{-1}$, however, and therefore a small colour correction must be applied 
to the blue sources (cf. Sect.~\ref{excess} for blue vs. red). This 
correction is obtained by dividing the above conversion factors by 1.05 and 
1.02 for LW2 and LW3, respectively, yielding the effective factors 2.21 and 
1.92 ADU/gain/s/mJy for blue sources.

Conversion from flux density to magnitude is defined as m$_{\rm 6.7} = -2.5 
\log (F_{\nu}(6.7 \mu m)/82.8)$ and m$_{\rm 14.3 \mu m} = -2.5 \log 
(F_{\nu}(14.3)/18.9)$, where $F_{\nu}$ is given in Jy. The $\sim 5$ \% 
responsivity decrease throughout orbit has not been corrected for. 

\subsection{Nordic Optical Telescope near-IR imaging}

A $6\arcmin \times 8\arcmin$ region inside Serp-CE, the area which is
usually referred to as the Serpens Cloud Core and has been covered to 
a smaller or larger extent by several studies in the near-IR
\citep{eir92,sog97,gio98,kaa99a}, was mapped deeply in $J$ (1.25 $\mu$m), 
$H$ (1.65 $\mu$m) and $K$ (2.2 $\mu$m) in August 1996, only 4 months 
after the ISOCAM observations. See Fig.~\ref{figmap} for the location
of the different maps. Also, a region to the NW of the $JHK$ field has 
been mapped in the $K$ band, see Fig.~\ref{fig-nw}, but with a total 
coadded integration time from only 30 sec to 1 min. 
These observations were made with the ARcetri Near Infrared CAmera 
(Arnica) at the 2.56m Nordic Optical Telescope, La Palma. See 
\citet{kaa99a} for details on this near-IR dataset.

\subsection{IRAM 30m Telescope Observations}

A 1.3~mm dust continuum mosaic of the Serpens main cloud core was taken with 
the IRAM 30-m telescope equipped with the MPIfR 37-channel bolometer array 
MAMBO-I \citep{kre98} during four night observing sessions in March 1998. 
The passband of the MAMBO bolometer array has an equivalent width $\approx 
70$~GHz and is centered at $\nu_{eff} \approx 240$~GHz. 

The $\sim 19' \times 6'$ mosaic consists of eleven individual on-the-fly maps 
which were obtained in the dual-beam raster mode with a scanning velocity of 
8$\arcsec$/sec and a spatial sampling of 4$\arcsec$ in elevation. In this mode, 
the telescope is scanning continuously in azimuth along each mapped row while 
the secondary mirror is wobbling in azimuth at frequency of 2~Hz. A wobbler 
throw of 45$\arcsec$ or 60$\arcsec$ was used. The typical azimuthal size of 
individual maps was 4$\arcmin$. The size of the main beam was measured to be 
$\sim$~11$\arcsec$ (HPBW) on Uranus and other strong point-like sources such 
as quasars. The pointing of the telescope was checked every $\sim 1$~hr 
using the VLA position of the strong, compact Class~0 source FIRS1 
(good to $\sim 0.1\arcsec $ -- \citet{cur93}); it was found to be accurate to 
better than $\sim $~3$\arcsec$. The zenith atmospheric optical depth, monitored 
by `skydips' every $\sim 2$~hr, was between $\sim 0.2$ and $\sim 0.4$. 
Calibration was achieved through on-the-fly mapping and on-off observations
of the primary calibrator Uranus \citep[e.g.][and references therein]{gri93}.
In addition, the Serpens secondary calibrator FIRS1, which has a 1.3~mm peak 
flux density $\sim 2.4$~Jy in an 11$\arcsec$ beam was observed before and after 
each map. The relative calibration was found to be good to within $\sim 10\%$ 
by comparing the individual coverages of each field, while the overall 
absolute calibration uncertainty is estimated to be $\sim 20\%$. 

The dual-beam maps were reduced and combined with the IRAM software for 
bolometer-array data (``NIC''; cf. \citet{bro95}) which uses the EKH 
restoration algorithm \citep{eme79}.  



\section{ISOCAM results}
\label{results}

\subsection{Source statistics, sensitivity and completeness}
\label{stat}

Table~\ref{tbl-1} gives an overview of the observational parameters and 
the results of the point source photometry for each of the six rasters 
named in col. 1. Columns 2, 3 and 4 give the centre position and size 
of the raster fields. Columns 5, 6 and 7 give the unit integration times, 
the PFOV and the average number of readouts per sky position. Column 8 
gives the total number of source detections. Columns 9 and 10 give the
number of sources for which photometry was obtained at 6.7 and 14.3 
$\mu$m, respectively. Column 11 gives the number of sources with flux 
determinations in both of the photometric bands. Columns 12 and 13 give 
the measured 1$\sigma$ photometric limits which are 
about 4 and 6 times the read-out-noise 
for the large and the deep fields, respectively. Practically all 
sources detected at 14.3 $\mu$m are also detected at 6.7 $\mu$m, but there 
are 15 cases of 14.3 $\mu$m detections without 6.7 $\mu$m counterparts. 
On the average, only 30\% of the 6.7 $\mu$m detections are also detected 
at 14.3 $\mu$m, but this value is clearly larger for the CE field. The
sensitivity, which depends on the PFOV and T$_{\rm int}$ as well as the amount 
of nebulosity and bright point sources, is lowered by memory effects 
from bright sources in the CE field and by the presence of a nebula in 
the CS field. For the deep fields (D1, D2, D3) the main limiting factor in 
terms of sensitivity is the memory effect and the fact that the flat fields 
are based on fewer frames. 

The deep fields within the CE area are independently repeated observations 
(observed within the same satellite orbit), and therefore permit a check 
of the photometric repeatability for a number of sources. The median values 
of the individual direct scatter between the measured fluxes were found to 
be: about 30\% for the 6.7 $\mu$m band measured from 28 sources between 4 mJy 
and 2 Jy, and about 15\% for the 14.3 $\mu$m band, measured from 14 sources 
between 30 mJy and 5.7 Jy (note different pixel scale). This corresponds to 
a median error of 0.13 dex in the $[14.3/6.7]$ colour index, which is in good 
agreement with the scatter (of sources without mid-IR excesses) around the 
value expected for normal photospheres (see Fig.~\ref{fig-323}) at a flux density 
of 10 mJy in the 14.3 $\mu$m band. 

To estimate how many field stars to expect at a given sensitivity
within the observed fields, the stellar content in a cone along the 
line-of-sight was integrated in steps of 0.02 kpc out to a distance 
of 20 kpc. The Galaxy was represented by an exponential disk, a bulge, 
a halo and a molecular ring, following \citet{wai92}.
Absolute magnitudes at 2.2 $\mu$m (M$_{\rm K}$) and in the 12 $\mu$m 
IRAS band (M$_{\rm 12}$), local number densities and scale heights of 
the different stellar populations, as well as their contribution to the 
different components of the Galaxy were provided by \citet{wai92} in their 
model of the mid-IR point source sky. 
Absolute magnitudes at 7 $\mu$m, M$_{\rm 7}$, are estimated by linear 
interpolation between M$_K$ and M$_{\rm 12}$.


\begin{figure}[htbp]
\resizebox{\hsize}{!}{\includegraphics{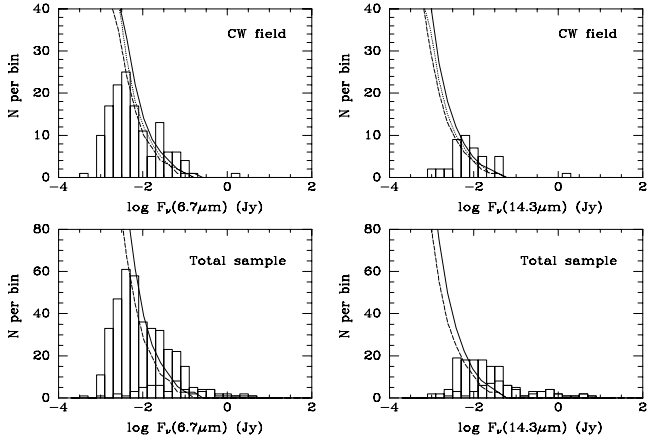}}
\caption{Histograms of sources at 6.7 $\mu$m (left) and 14.3 $\mu$m (right) 
         for the CW field (upper) and the total sample (lower). The boldface 
         steps in the lower panels correspond to the population of mid-IR 
         excess sources. The expected number of galactic sources in each bin, 
         normalised to each field, from modelling the stellar content in a 
         cone along the line-of-sight is inserted for zero cloud extinction 
         (solid lines), A$_V \approx 5$ (dotted lines), and A$_V \approx 10$ 
         (dashed lines). \label{fig1}}
\end{figure}

The upper panels of Fig.~\ref{fig1} show the histograms of the 6.7 $\mu$m 
and 14.3 $\mu$m sources in the CW field, which is free of nebulosity and 
practically free of IR excess sources. The bin size of 0.2 in $\log F_{\nu}$ 
corresponds to 0.5 magnitudes. Inserted are the model counts of galactic 
sources at 7 and 12 $\mu$m, scaled to the CW field size. The expected source 
number per bin is calculated assuming no cloud extinction (solid line), an 
average extinction of 0.2 magnitudes in the 6.7 $\mu$m band (dotted line), 
and A$_{\rm 6.7}$ = 0.4 magnitudes (dashed line), corresponding to roughly 
A$_V \sim 5$ and A$_V \sim 10$, respectively (cf. Sects.~\ref{nir} and 
\ref{char} about extinction).
For the CW field an average cloud extinction of A$_V \sim 7-8$ magnitudes 
is in agreement with the location of the majority of the sources in a 
DENIS $I-J/J-K'$ diagram, as well as an extinction map based on R star 
counts \citep{cam99}. 
Comparing the number of observed sources at 14.3 $\mu$m with the model 
expectation at 12 $\mu$m indicates completeness at $\sim$ 6 mJy or
m$_{\rm 14.3}$ = 8.7 mag. The observations at 6.7 $\mu$m are estimated
to be complete to $\sim $ 5 mJy or m$_{\rm 6.7}$ = 10.6 mag. 
The lower panels of Fig.~\ref{fig1} show the histograms for the total
sample. The boldface steps show the contribution of the mid-IR excess
sources. The dashed line represents the model expectation assuming an 
average cloud extinction of about 10 magnitudes of visual extinction,
i.e. A$_{\rm 6.7}$ = 0.41 mag and A$_{\rm 14.3}$ = 0.36 mag (cf. 
Sects.~\ref{nir} and \ref{char}). 
Thus, the whole sample taken together, and allowing for a subtraction
of the mid-IR excess sources, indicates an overall completeness at 
$\sim$ 6 mJy for the 6.7 $\mu$m band and at 8 mJy for the 14.3 $\mu$m 
band (i.e. at m$_{\rm 6.7}$ = 10.4 mag and m$_{\rm 14.3}$ = 8.5 mag).


\begin{figure}[htbp]
\resizebox{\hsize}{!}{\includegraphics{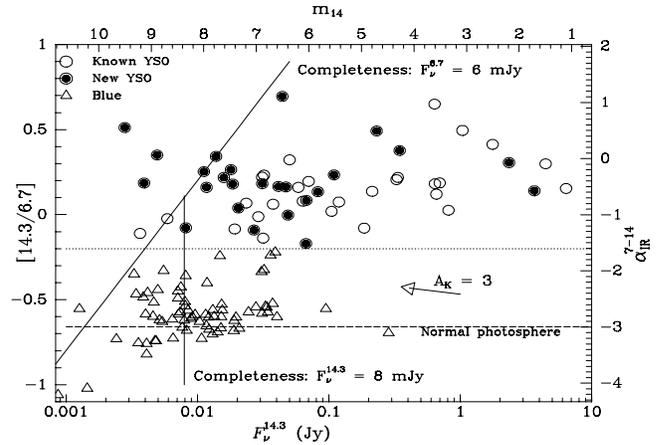}}
\caption{The colour index $[14.3/6.7]$ defined as $\log 
         (F_{\nu}^{14.3}/F_{\nu}^{6.7})$ is shown on the y-axis 
         and the flux $F_{\nu}^{14.3}$ also given in magnitudes 
         $m_{14}$ is shown on the x-axis. The SED index between 6.7 and 
         14.3 $\mu$m is given on the right hand y-axis. 53 sources with 
         substantial IR excess are located above $[14.3/6.7] = -0.2$ or 
         $\alpha_{\rm IR}^{7-14} = -1.6$ (dotted line). While 28 of these 
         were previously suggested as YSOs, 25 are new YSO candidates 
         (filled circles). \label{fig-323}}
\end{figure}  

\subsection{Sources with mid-IR excesses}
\label{excess}

For the 124 sources with fluxes in both bands we present a colour 
magnitude diagram in Fig.~\ref{fig-323}. The colour index $[14.3/6.7]$, 
defined as $ \log (F_{\nu}^{14.3}/F_{\nu}^{6.7})$, is plotted against 
$F_{\nu}^{14.3}$. This colour index can be converted to the commonly used
index of the Spectral Energy Distribution (SED):
\begin{equation}
\alpha_{\rm IR}^{7-14} = \frac{\log (\lambda_{14} F_{\lambda_{14}}) 
                             - \log (\lambda_7 F_{\lambda_7})}
                {\log \lambda_{14} - \log \lambda_7}, 
\end{equation}
calculated between 6.7 and 14.3 $\mu$m and indicated on the right hand 
y-axis. The completeness limit is given with solid lines as the combined 
effect of the completeness in each of the two bands. 

The sources tend to separate into two distinct groups, with some few
intermediate objects. The ``red'' sources (circles) are interpreted as 
pre-main-sequence (PMS) stars surrounded by circumstellar dust. This
sample is not believed to be contaminated with galaxies. At the 
completeness flux level of 8 mJy at 14.3 $\mu$m the expected extracalagtic 
contamination is about half a source within our map coverage and at the 
level of 3 mJy (the faintest source in our sample at 14.3 $\mu$m) the 
contamination is below two sources within the mapped region \citep{hon03}.
No correction for extinction has been made on these numbers, so that they should be
considered as upper limits to the extragalactic contamination.

We have decided to set the division line between 
mid-IR excess objects and ``blue'' objects (triangles) at $[14.3/6.7] = 
-0.2$ or $\alpha_{\rm IR}^{7-14} = -1.6$ (dotted line), which corresponds 
to the classical border between Class~II and Class~III objects 
(cf Sect~\ref{char}). 

\begin{table*}

\caption{The 77 ISOCAM sources identified as YSOs in the Serpens cluster. 
         \label{tbl-2}
         }

\newcommand\cola {\null}
\newcommand\colc {&}
\newcommand\cole {&}
\newcommand\colh {&}
\newcommand\coli {&}
\newcommand\colj {&}
\newcommand\colk {&}
\newcommand\coll {&}
\newcommand\colm {&}
\newcommand\coln {&}
\newcommand\colo {&}
\newcommand\colp {&}
\newcommand\colq {&}
\newcommand\colr {&}
\newcommand\eol{\\}
\newcommand\extline{&&&&&&&&&&&&&\eol}

\[
   \begin{tabular}{rccrrrrrrrrrrl}
   \hline
   \noalign{\smallskip}

{ISO} & $\alpha_{2000}$ & $\delta_{2000}$ & {$J$} & {$\sigma_{J}$} & {$H$} & $\sigma_{H}$ & {$K$} & $\sigma_{K}$ & $F_{\nu}^{6.7}$ & $\sigma_{6.7}$ & $F_{\nu}^{14.3}$ & $\sigma_{14.3}$ & {Other ID} \\
 \# & 18$^h$ &  & mag & & mag & & mag &  & mJy & mJy & mJy & mJy & \\
\hline
\cola  29\colc  28$^m$52$^s_.$5\cole 1$^{\circ}$12$\arcmin$47$\arcsec$\colh \coli \colj 
\colk \coll \colm \coln    1$^g$\colo   0.6\colp    3\colq   2\colr \eol
\cola 150\colc  29$^m$30$^s_.$6\cole 1$^{\circ}$01$\arcmin$09$\arcsec$\colh \coli \colj 
\colk \coll \colm \coln   19\colo   3\colp   21\colq   5\colr \eol
\cola 158\colc  29$^m$31$^s_.$7\cole  1$^{\circ}$08$\arcmin$22$\arcsec$\colh \coli \colj 
\colk \coll \colm \coln    10\colo   1\colp   16$^m$\colq   7\colr BD+01 3687\eol
\cola 159\colc  29$^m$32$^s_.$0\cole  1$^{\circ}$18$\arcmin$42$\arcsec$\colh \coli \colj
\colk \coll  8.41\colm $<$0.01\coln 1162\colo  28\colp 2351\colq  26\colr IRAS 18269+0116\eol
\cola 160\colc 29$^m$32$^s_.$0\cole 1$^{\circ}$18$\arcmin$34$\arcsec$\colh \coli \colj 
\colk \coll 9.82\colm $<$0.01\coln 114\colo 2\colp -\colq -\colr IRAS 18269+0116\eol
\cola 173\colc  29$^m$33$^s_.$4\cole  1$^{\circ}$08$\arcmin$28$\arcsec$\colh \coli \colj 
\colk \coll \colm \coln   64\colo   2\colp  110\colq  20\colr [CDF88] 6\eol
\cola 202\colc  29$^m$39$^s_.$9\cole  1$^{\circ}$17$\arcmin$55$\arcsec$\colh \coli \colj 
\colk \coll 11.80\colm $<$0.01\coln    8\colo   2\colp   12$^c$\colq   3\colr \eol
\cola 207\colc  29$^m$41$^s_.$1\cole   1$^{\circ}$07$\arcmin$40$\arcsec$\colh \coli \colj 
\colk \coll \colm \coln   56\colo   1\colp   68\colq   3\colr STGM3\eol
\cola 216\colc  29$^m$42$^s_.$3\cole     1$^{\circ}$20$\arcmin$19$\arcsec$\colh \coli \colj 
\colk \coll 12.02\colm $<$0.01\coln    6\colo   3\colp   14$^c$\colq   2\colr \eol
\cola 219\colc  29$^m$43$^s_.$7\cole      1$^{\circ}$07$\arcmin$22$\arcsec$\colh \coli \colj 
\colk \coll \colm \coln    7\colo   2\colp  -\colq -\colr STGM2\eol
\cola 221\colc  29$^m$44$^s_.$3\cole       1$^{\circ}$04$\arcmin$55$\arcsec$\colh \coli \colj 
\colk \coll \colm \coln 2652\colo  37\colp 3664\colq  23\colr IRAS 18271+0102\eol
\cola 224\colc  29$^m$44$^s_.$8\cole      1$^{\circ}$13$\arcmin$10$\arcsec$\colh 13.22\coli $<$0.01\colj 12.26
\colk $<$0.01\coll 11.82\colm $<$0.01\coln    7\colo   2\colp  -\colq -\colr EC11\eol
\cola 226\colc  29$^m$44$^s_.$8\cole      1$^{\circ}$15$\arcmin$44$\arcsec$\colh -\coli -\colj 15.45
\colk 0.02\coll 13.51\colm $<$0.01\coln    3\colo 1\colp  -\colq -\colr EC13\eol
\cola 231\colc  29$^m$46$^s_.$0\cole      1$^{\circ}$16$\arcmin$23$\arcsec$\colh 18.12\coli 0.09\colj 13.86
\colk $<$0.01\coll 11.81\colm $<$0.01\coln   11\colo 4\colp  -\colq -\colr EC21\eol
\cola 232\colc  29$^m$46$^s_.$3\cole      1$^{\circ}$12$\arcmin$14$\arcsec$\colh 13.22\coli $<$0.01\colj 11.66
\colk $<$0.01\coll 10.99\colm $<$0.01\coln    6\colo   4\colp  -\colq -\colr EC23\eol
\cola 234\colc  29$^m$46$^s_.$9\cole      1$^{\circ}$16$\arcmin$10$\arcsec$\colh -\coli -\colj 18.73
\colk 0.14\coll 14.47\colm 0.02\coln    7\colo 3\colp  -\colq -\colr EC26\eol
\cola 237\colc  29$^m$47$^s_.$2\cole      1$^{\circ}$16$\arcmin$26$\arcsec$\colh -\coli -\colj 16.51
\colk 0.03\coll 13.19\colm 0.01\coln   33\colo   3\colp   27\colq   5\colr EC28\eol
\cola 241\colc  29$^m$48$^s_.$1\cole      1$^{\circ}$16$\arcmin$43$\arcsec$\colh -\coli -\colj -
\colk -\coll 16.11\colm 0.05\coln   28\colo   4\colp   41\colq   8\colr  \eol
\cola 242\colc  29$^m$48$^s_.$6\cole      1$^{\circ}$13$\arcmin$42$\arcsec$\colh 18.14\coli 0.10\colj 14.94
\colk $<$0.01\coll 13.14\colm 0.01\coln    5\colo   2\colp    4\colq   1\colr K8,EC33\eol
\cola 249\colc  29$^m$49$^s_.$1\cole      1$^{\circ}$16$\arcmin$32$\arcsec$\colh -\coli -\colj 17.09
\colk 0.05\coll 13.62\colm 0.01\coln  143$^f$\colo   6\colp  637$^f$\colq   9\colr EC37\eol
\cola 250\colc  29$^m$49$^s_.$3\cole      1$^{\circ}$16$\arcmin$19$\arcsec$\colh -\coli -\colj -
\colk -\coll 11.67\colm $<$0.01\coln 2242$^f$\colo  54\colp 4464$^f$\colq  35\colr DEOS\eol
\cola 252\colc  29$^m$49$^s_.$9\cole   0$^{\circ}$56$\arcmin$12$\arcsec$\colh \coli \colj 
\colk \coll \colm \coln   21\colo   1\colp   31\colq   3\colr \eol
\cola 253\colc  29$^m$49$^s_.$6\cole      1$^{\circ}$14$\arcmin$57$\arcsec$\colh -\coli -\colj -
\colk -\coll 14.90\colm 0.03\coln   49\colo  13\colp   49\colq   5\colr EC40\eol
\cola 254\colc  29$^m$49$^s_.$5\cole      1$^{\circ}$17$\arcmin$07$\arcsec$\colh \coli \colj 
\colk \coll 12.46\colm 0.01\coln  156\colo  12\colp  214\colq   5\colr EC38\eol
\cola 258a\colc  29$^m$49$^s_.$6\cole    1$^{\circ}$15$\arcmin$28$\arcsec$\colh -\coli -\colj 17.77
\colk 0.08\coll 14.80\colm 0.03\coln  19$^f$\colo 2\colp 24$^f$\colq 4\colr EC41,GCNM23\eol
\cola 258b\colc  29$^m$50$^s_.$3\cole    1$^{\circ}$15$\arcmin$21$\arcsec$\colh -\coli -\colj
-\colk -\coll -\colm -\coln  -\colo   -\colp   13$^f$\colq   4\colr \eol
\cola 259\colc  29$^m$50$^s_.$6\cole       1$^{\circ}$01$\arcmin$35$\arcsec$\colh \coli \colj 
\colk \coll \colm \coln   29\colo  13\colp  -\colq -\colr \eol
\cola 260\colc  29$^m$50$^s_.$5\cole      1$^{\circ}$14$\arcmin$17$\arcsec$\colh -\coli -\colj -
\colk -\coll 16.76\colm 0.15\coln    3\colo   1\colp  -\colq -\colr \eol
\cola 265\colc  29$^m$51$^s_.$2\cole      1$^{\circ}$16$\arcmin$42$\arcsec$\colh 15.35\coli 0.04\colj 12.90
\colk 0.01\coll 11.32\colm $<$0.01\coln  766\colo  13\colp  813\colq  16\colr EC53\eol
\cola 266\colc  29$^m$51$^s_.$3\cole      1$^{\circ}$13$\arcmin$17$\arcsec$\colh 15.90\coli 0.02\colj 14.27
\colk 0.01\coll 13.02\colm $<$0.01\coln    2\colo   1\colp  -\colq -\colr GCNM35,EC51\eol
\cola 269\colc  29$^m$52$^s_.$2\cole      1$^{\circ}$13$\arcmin$21$\arcsec$\colh -\coli -\colj 14.79
\colk 0.01\coll 13.31\colm 0.01\coln    2\colo   1\colp  -\colq -\colr K16,EC56\eol
\cola 270\colc  29$^m$52$^s_.$2\cole      1$^{\circ}$15$\arcmin$49$\arcsec$\colh -\coli -\colj -
\colk -\coll 17.70\colm 0.20\coln    9\colo   2\colp   44\colq   8\colr \eol
\cola 272\colc  29$^m$52$^s_.$5\cole      1$^{\circ}$12$\arcmin$54$\arcsec$\colh 19.01\coli 0.22\colj 14.63
\colk 0.01\coll 12.54\colm $<$0.01\coln    3\colo   1\colp  -\colq -\colr EC59\eol
\cola 276\colc  29$^m$52$^s_.$9\cole      1$^{\circ}$14$\arcmin$56$\arcsec$\colh -\coli -\colj -
\colk -\coll 14.80\colm 0.03\coln   74\colo  12\colp  231\colq   4\colr GCNM53\eol
\cola 277\colc  29$^m$53$^s_.$2\cole      1$^{\circ}$15$\arcmin$43$\arcsec$\colh -\coli -\colj -
\colk -\coll 14.88\colm 0.03\coln    9\colo   2\colp  -\colq -\colr EC63\eol
\cola 279\colc  29$^m$53$^s_.$4\cole      1$^{\circ}$13$\arcmin$13$\arcsec$\colh -\coli -\colj 14.15
\colk 0.01\coll 13.60\colm 0.01\coln    1\colo   2\colp  -\colq -\colr STGM14,EC66\eol
\cola 283\colc  29$^m$53$^s_.$6\cole      1$^{\circ}$17$\arcmin$00$\arcsec$\colh -\coli -\colj 11.58$^a$
\colk $<$0.01\coll 10.63$^a$\colm $<$0.01\coln   30\colo   4\colp   29\colq   6\colr EC67\eol
\cola 285\colc  29$^m$53$^s_.$9\cole      1$^{\circ}$13$\arcmin$32$\arcsec$\colh 16.58\coli 0.04\colj 14.21
\colk 0.01\coll 12.87\colm $<$0.01\coln    6$^d$\colo   1\colp    6\colq   3\colr GCNM63,EC68\eol
\cola 287\colc  29$^m$54$^s_.$1\cole      1$^{\circ}$07$\arcmin$14$\arcsec$\colh \coli \colj 
\colk \coll \colm \coln    3\colo   1\colp    4\colq   3\colr \eol
\cola 289\colc  29$^m$54$^s_.$4\cole      1$^{\circ}$15$\arcmin$03$\arcsec$\colh 16.15\coli 0.06\colj 12.66
\colk $<$0.01\coll 10.86\colm $<$0.01\coln   19\colo   9\colp  -\colq -\colr EC69,CK10\eol
\cola 291\colc  29$^m$54$^s_.$4\cole      1$^{\circ}$14$\arcmin$44$\arcsec$\colh 14.26\coli 0.01\colj 13.32
\colk $<$0.01\coll 12.74\colm $<$0.01\coln    2\colo   2\colp  -\colq -\colr GCNM70,EC70\eol
\cola 294\colc  29$^m$55$^s_.$1\cole      1$^{\circ}$13$\arcmin$22$\arcsec$\colh 16.32\coli 0.03\colj 13.84
\colk $<$0.01\coll 12.39\colm $<$0.01\coln   19\colo   3\colp   31\colq  12\colr EC73,GEL3\eol
\cola 298\colc  29$^m$55$^s_.$6\cole      1$^{\circ}$14$\arcmin$31$\arcsec$\colh 14.78\coli 0.02\colj 12.21
\colk $<$0.01\coll 10.44\colm $<$0.01\coln  100\colo  27\colp  119$^d$\colq  16\colr EC74,CK9,GEL4\eol
\cola 304\colc  29$^m$56$^s_.$6\cole      1$^{\circ}$13$\arcmin$01$\arcsec$\colh 14.86\coli 0.01\colj 12.55
\colk $<$0.01\coll 11.38\colm $<$0.01\coln   33\colo   5\colp   38\colq   3\colr EC79,GEL5\eol
\cola 306\colc  29$^m$56$^s_.$6\cole      1$^{\circ}$12$\arcmin$40$\arcsec$\colh -\coli -\colj 17.90
\colk 0.08\coll 13.75\colm 0.02\coln   43\colo   4\colp   32\colq   4\colr EC80\eol
\cola 307\colc  29$^m$56$^s_.$8\cole      1$^{\circ}$14$\arcmin$46$\arcsec$\colh 12.01\coli $<$0.01\colj 10.20
\colk $<$0.01\coll  8.63\colm $<$0.01\coln  680$^{e,d}$\colo  10\colp 1762$^{e,d}$\colq  43\colr SVS2,CK3,EC82\eol
\cola 308\colc  29$^m$56$^s_.$8\cole      1$^{\circ}$13$\arcmin$18$\arcsec$\colh -\coli -\colj -
\colk -\coll 16.62\colm 0.19\coln   11.0\colo   2\colp  -\colq -\colr K32,HCE170/171\eol
\cola 309\colc  29$^m$56$^s_.$8\cole      1$^{\circ}$12$\arcmin$49$\arcsec$\colh 15.15\coli 0.02\colj 12.44
\colk $<$0.01\coll 11.00\colm $<$0.01\coln   45\colo   4\colp   70\colq   3\colr EC84,GEL7\eol
\cola 312\colc  29$^m$57$^s_.$5\cole      1$^{\circ}$12$\arcmin$59$\arcsec$\colh 15.56$^b$\coli 0.03\colj 13.61$^b$
\colk 0.01\coll 11.24$^b$\colm $<$0.01\coln  453\colo   8\colp  697\colq  10\colr EC88+EC89\eol
\cola 313\colc  29$^m$57$^s_.$5\cole      1$^{\circ}$13$\arcmin$49$\arcsec$\colh 17.81\coli 0.12\colj 15.47
\colk 0.04\coll 13.46\colm 0.03\coln   32\colo   3\colp  -\colq -\colr GCNM94,EC87\eol
\cola 314\colc  29$^m$57$^s_.$5\cole      1$^{\circ}$14$\arcmin$07$\arcsec$\colh 12.10\coli $<$0.01\colj  9.16
\colk $<$0.01\coll  7.03\colm $<$0.01\coln 4479\colo  97\colp 6388\colq  50\colr SVS20,CK1,EC90\eol
\cola 317\colc  29$^m$57$^s_.$8\cole      1$^{\circ}$12$\arcmin$52$\arcsec$\colh 15.30$^b$\coli 0.02\colj 11.47$^b$
\colk $<$0.01\coll  9.45$^b$\colm $<$0.01\coln  332\colo   5\colp 1039$^d$\colq  35\colr EC92+EC95\eol
\cola 318\colc  29$^m$57$^s_.$8\cole      1$^{\circ}$12$\arcmin$37$\arcsec$\colh -\coli -\colj 14.22
\colk 0.01\coll 11.47\colm $<$0.01\coln  100$^f$\colo   9\colp  105$^f$\colq  20\colr EC94\eol
\cola 319\colc  29$^m$57$^s_.$7\cole      1$^{\circ}$15$\arcmin$31$\arcsec$\colh 15.95\coli 0.03\colj 12.67
\colk $<$0.01\coll 10.59\colm $<$0.01\coln   45\colo   4\colp  -\colq -\colr EC93,CK13\eol
\cola 320\colc  29$^m$57$^s_.$8\cole      1$^{\circ}$12$\arcmin$29$\arcsec$\colh -\coli -\colj 15.61
\colk 0.02\coll 12.91\colm $<$0.01\coln   23\colo   7\colp   19\colq   5\colr EC91\eol
\cola 321\colc  29$^m$58$^s_.$2\cole      1$^{\circ}$15$\arcmin$22$\arcsec$\colh 12.97\coli $<$0.01\colj 10.92
\colk $<$0.01\coll  9.48\colm $<$0.01\coln  204\colo   9\colp  327$^d$\colq   5\colr CK4,GEL12,EC97\eol
\cola 322\colc  29$^m$58$^s_.$3\cole      1$^{\circ}$12$\arcmin$49$\arcsec$\colh -\coli -\colj 14.61
\colk 0.01\coll 12.25\colm $<$0.01\coln   32\colo   3\colp  -\colq -\colr EC98\eol
\cola 326\colc  29$^m$58$^s_.$7\cole      1$^{\circ}$14$\arcmin$26$\arcsec$\colh 16.62\coli 0.06\colj 13.82
\colk 0.01\coll 11.56\colm $<$0.01\coln  202\colo  39\colp  335$^f$\colq   8\colr EC103\eol
   \noalign{\smallskip}
   \hline
   \end{tabular}
\]

         {\small (continued on next page)} \\
         
\end{table*}

\begin{table*}

\newcommand\cola {\null}
\newcommand\colc {&}
\newcommand\cole {&}
\newcommand\colh {&}
\newcommand\coli {&}
\newcommand\colj {&}
\newcommand\colk {&}
\newcommand\coll {&}
\newcommand\colm {&}
\newcommand\coln {&}
\newcommand\colo {&}
\newcommand\colp {&}
\newcommand\colq {&}
\newcommand\colr {&}
\newcommand\eol{\\}
\newcommand\extline{&&&&&&&&&&&&&\eol}

\[
   \begin{tabular}{rccrrrrrrrrrrl}
   \hline
   \noalign{\smallskip}

{ISO} & $\alpha_{2000}$ & $\delta_{2000}$ & {$J$} & {$\sigma_{J}$} & {$H$} & $\sigma_{H}$ & {$K$} & $\sigma_{K}$ & 
$F_{\nu}^{6.7}$ & $\sigma_{6.7}$ & $F_{\nu}^{14.3}$ & $\sigma_{14.3}$ & {Other ID} \\
 \# & 18$^h$ &  & mag & & mag & & mag & & mJy & mJy & mJy & mJy & \\
\hline
\cola 327\colc  29$^m$58$^s_.$9\cole      1$^{\circ}$12$\arcmin$31$\arcsec$\colh -\coli -\colj -
\colk -\coll 15.41\colm 0.04\coln    9.7\colo   2.0\colp    8.1\colq   4.3\colr HCE175\eol
\cola 328\colc  29$^m$59$^s_.$2\cole      1$^{\circ}$14$\arcmin$06$\arcsec$\colh 11.78\coli $<$0.01\colj 10.36
\colk $<$0.01\coll  9.46\colm $<$0.01\coln  223\colo  23\colp  186$^f$\colq  10\colr EC105,CK8,GEL13\eol
\cola 330\colc  29$^m$59$^s_.$5\cole      1$^{\circ}$11$\arcmin$59$\arcsec$\colh -\coli -\colj -
\colk -\coll 14.37\colm 0.02\coln  423\colo   7\colp  643\colq   9\colr HB1\eol
\cola 331\colc  29$^m$59$^s_.$7\cole      1$^{\circ}$13$\arcmin$13$\arcsec$\colh -\coli -\colj -
\colk -\coll 15.37\colm 0.04\coln  145\colo   9\colp  346\colq  12\colr \eol
\cola 338\colc  30$^m$00$^s_.$7\cole      1$^{\circ}$13$\arcmin$38$\arcsec$\colh 13.40\coli $<$0.01\colj 11.15
\colk $<$0.01\coll 10.07\colm $<$0.01\coln   16\colo   4\colp  -\colq -\colr EC117,CK6,GEL15\eol
\cola 341\colc  30$^m$01$^s_.$1\cole      1$^{\circ}$13$\arcmin$26$\arcsec$\colh -\coli -\colj 15.39
\colk 0.02\coll 12.90\colm $<$0.01\coln   20\colo   3\colp   24\colq  11\colr EC121\eol
\cola 345\colc  30$^m$02$^s_.$1\cole      1$^{\circ}$14$\arcmin$00$\arcsec$\colh 17.16\coli 0.06\colj 14.74
\colk 0.01\coll 12.88\colm $<$0.01\coln   24\colo   4\colp   50\colq   7\colr EC125,CK7\eol
\cola 347\colc  30$^m$02$^s_.$8\cole      1$^{\circ}$12$\arcmin$28$\arcsec$\colh 15.11\coli 0.02\colj 11.72
\colk $<$0.01\coll  9.69\colm $<$0.01\coln  499$^c$\colo  12\colp  658\colq  13\colr EC129\eol
\cola 348\colc  30$^m$03$^s_.$2\cole      1$^{\circ}$16$\arcmin$17$\arcsec$\colh 12.14\coli $<$0.01\colj 10.89
\colk $<$0.01\coll 10.15\colm $<$0.01\coln   41$^n$\colo  11\colp   59$^d$\colq   3\colr EC135,GGD29\eol
\cola 351\colc  30$^m$04$^s_.$0\cole      1$^{\circ}$12$\arcmin$38$\arcsec$\colh 16.51\coli 0.03\colj 13.54
\colk $<$0.01\coll 12.13\colm $<$0.01\coln    8.4\colo   1.9\colp  -\colq -\colr EC141\eol
\cola 356\colc  30$^m$05$^s_.$2\cole      1$^{\circ}$12$\arcmin$44$\arcsec$\colh 16.08\coli 0.03\colj 14.51
\colk 0.01\coll 13.53\colm 0.01\coln    2.5\colo   1.6\colp  -\colq -\colr K40,EC152\eol
\cola 357\colc  30$^m$05$^s_.$8\cole      1$^{\circ}$06$\arcmin$22$\arcsec$\colh \coli \colj 
\colk \coll \colm \coln    10\colo   1.7\colp   18\colq   2.3\colr \eol
\cola 359\colc  30$^m$06$^s_.$4\cole      1$^{\circ}$01$\arcmin$09$\arcsec$\colh \coli \colj 
\colk \coll \colm \coln   32\colo   4\colp   47\colq   4\colr \eol
\cola 366\colc  30$^m$07$^s_.$6\cole      1$^{\circ}$12$\arcmin$04$\arcsec$\colh 12.20\coli $<$0.01\colj 10.74
\colk $<$0.01\coll  9.92\colm $<$0.01\coln   53\colo   7\colp   64\colq   5\colr STGM8\eol
\cola 367\colc  30$^m$08$^s_.$1\cole      1$^{\circ}$01$\arcmin$42$\arcsec$\colh \coli \colj 
\colk \coll \colm \coln   12$^n$\colo   5\colp   19$^n$\colq   4\colr \eol
\cola 370\colc  30$^m$08$^s_.$4\cole      0$^{\circ}$58$\arcmin$48$\arcsec$\colh \coli \colj 
\colk \coll \colm \coln   61\colo   3\colp   83\colq   4\colr double in I\eol
\cola 379\colc  30$^m$09$^s_.$3\cole      1$^{\circ}$02$\arcmin$47$\arcsec$\colh \coli \colj 
\colk \coll \colm \coln   99\colo   5\colp   67\colq   6\colr \eol
\cola 393\colc  30$^m$11$^s_.$2\cole      1$^{\circ}$12$\arcmin$40$\arcsec$\colh 12.97$^b$\coli $<$0.01\colj 11.67$^b$
\colk $<$0.01\coll 11.06$^b$\colm $<$0.01\coln    6.3\colo   4.1\colp   11.2\colq   4.5\colr \eol
\cola 407\colc  30$^m$13$^s_.$9\cole      1$^{\circ}$08$\arcmin$55$\arcsec$\colh \coli \colj 
\colk \coll \colm \coln    2.2\colo   1.3\colp    4.9\colq   5.3\colr \eol
   \noalign{\smallskip}
   \hline
   \end{tabular}     
\]

$^{a}$ $H$ and $K$ band data from the Arnica 1995 map, which is slightly 
       displaced from the 1996 map and therefore includes this object. \\
$^{b}$ ISOCAM source is resolved into two sources in the near-IR, and the
       near-IR fluxes are added.  \\
$^{c}$ Flux measurement affected by proximity to the dead column. If 
       the dead column cannot be avoided by any of the redundant observations, 
       this flagging. If source is located on the dead column, a flux 
       measurement is not attempted at all. \\   
$^{d}$ Source close to the detector edge in all redundant observations. \\
$^{e}$ Extended source. \\
$^{f}$ ISOCAM source is not quite resolved from a bright neighbour. \\
$^{g}$ Galaxy contamination? For fluxes $\sim$ 3 mJy at 14.3 $\mu$m  the 
       ``red'' sample is expected to contain less than two galaxies in our
       field according to \citet{hon03}. \\
$^{m}$ Flux might be affected by memory effects from other sources. \\
$^{n}$ Nebulous sky background. \\

   Empty space means no measurement available in this work, while a hyphen
   means no detection. \\

    Identifier acronyms are related to the following references: 
    SVS (Strom et al. 1976); GGD (Gyulbudaghian et al. 1978); HL (Hartigan \& 
    Lada 1985); CK (Churchwell \& Koornneef 1986); GEL (G\'omez de Castro et 
    al. 1988); CDF (Chavarria-K et al. 1988); EC (Eiroa \& Casali 1992); SMM 
    (Casali et al. 1993); MMW (McMullin et al. 1994); HHR (Hodapp et al. 1996); 
    HB (Hurt \& Barsony 1996); STGM (Sogawa et al. 1997); HCE (Horrobin et al. 
    1997); GCNM (Giovannetti et al. 1998); K (Kaas 1999)

\end{table*}


\begin{table*}
\caption{The 71 ``blue'' ISOCAM sources which have $[14.3/6.7] < -0.2$. Many 
         of these are likely to be field stars, but some could be young 
         cluster members of Class III type. Colour correction applied.
         \label{tbl-3}
         }

\newcommand\cola {\null}
\newcommand\colc {&}
\newcommand\cole {&}
\newcommand\colh {&}
\newcommand\coli {&}
\newcommand\colj {&}
\newcommand\colk {&}
\newcommand\coll {&}
\newcommand\colm {&}
\newcommand\coln {&}
\newcommand\colo {&}
\newcommand\colp {&}
\newcommand\colq {&}
\newcommand\colr {&}
\newcommand\eol{\\}
\newcommand\extline{&&&&&&&&&&&&&\eol}

\[
  \begin{tabular}{rccrrrrrrrrrrl}
  \hline
  \noalign{\smallskip} 

{ISO} & $\alpha_{2000}$ & $\delta_{2000}$ & {$J$} & {$\sigma_{J}$} & {$H$} & $\sigma_{H}$ & {$K$} & 
$\sigma_{K}$ & $F_{\nu}^{6.7}$ & $\sigma_{6.7}$ & $F_{\nu}^{14.3}$ & $\sigma_{14.3}$ & {Other ID} \\ 
\# & 18$^h$ &  & mag & & mag & & mag &  & mJy & mJy & mJy & mJy & \\
\hline
\cola   2\ \colc  28$^m$41.2\cole   1$^{\circ}$15$\arcmin$15$\arcsec$\colh  \coli  \colj  
\colk  \coll  \colm  \coln   30\colo   5\colp   12\colq   2\colr \eol
\cola   4\ \colc  28$^m$41.9\cole   1$^{\circ}$14$\arcmin$06$\arcsec$\colh  \coli  \colj  
\colk  \coll  \colm  \coln   35\colo   5\colp    6.4\colq   2.9\colr \eol
\cola   7\ \colc  28$^m$42.1\cole   1$^{\circ}$14$\arcmin$31$\arcsec$\colh  \coli  \colj  
\colk  \coll  \colm  \coln   26\colo   3\colp    7.9\colq   3.0\colr \eol
\cola   9\ \colc  28$^m$43.6\cole   1$^{\circ}$16$\arcmin$10$\arcsec$\colh  \coli  \colj  
\colk  \coll  \colm  \coln   19\colo   3\colp    8.0\colq   2.2\colr \eol
\cola  11\ \colc  28$^m$45.1\cole   1$^{\circ}$10$\arcmin$19$\arcsec$\colh  \coli  \colj  
\colk  \coll  \colm  \coln    10\colo   2\colp    3.3\colq   3.1\colr \eol
\cola  12\ \colc  28$^m$45.2\cole   1$^{\circ}$16$\arcmin$50$\arcsec$\colh  \coli  \colj  
\colk  \coll  \colm  \coln   57\colo   8\colp   11\colq   2\colr \eol
\cola  16\ \colc  28$^m$47.0\cole   1$^{\circ}$10$\arcmin$36$\arcsec$\colh  \coli  \colj  
\colk  \coll  \colm  \coln   53\colo   6\colp   13\colq   2\colr \eol
\cola  20\ \colc  28$^m$48.4\cole   1$^{\circ}$13$\arcmin$14$\arcsec$\colh  \coli  \colj  
\colk  \coll  \colm  \coln   37\colo   6\colp    9\colq   2\colr \eol
\cola  24\ \colc  28$^m$49.7\cole   1$^{\circ}$19$\arcmin$09$\arcsec$\colh  \coli  \colj  
\colk  \coll  \colm  \coln   60\colo   7\colp   15\colq   2\colr \eol
\cola  31\ \colc  28$^m$52.6\cole   1$^{\circ}$10$\arcmin$10$\arcsec$\colh  \coli  \colj  
\colk  \coll  \colm  \coln    5$^d$\colo   1\colp    1.2\colq   2.8\colr \eol
\cola  42\ \colc  28$^m$58.3\cole   1$^{\circ}$10$\arcmin$55$\arcsec$\colh  \coli  \colj  
\colk  \coll  \colm  \coln   20\colo   2\colp    3.5\colq   1.7\colr [SCB96] 41\eol
\cola  46\ \colc  28$^m$59.8\cole   1$^{\circ}$09$\arcmin$58$\arcsec$\colh  \coli  \colj  
\colk  \coll  \colm  \coln   27\colo   4\colp    4.8\colq   2.5\colr \eol
\cola  49\ \colc  29$^m$ 0.3\cole   1$^{\circ}$18$\arcmin$17$\arcsec$\colh  \coli  \colj  
\colk  \coll  \colm  \coln  114\colo  14\colp   32$^f$\colq   4\colr \eol
\cola  61\ \colc  29$^m$ 3.4\cole   1$^{\circ}$18$\arcmin$34$\arcsec$\colh  \coli  \colj  
\colk  \coll  \colm  \coln   28\colo   4\colp    8\colq   2\colr \eol
\cola  63\ \colc  29$^m$ 3.8\cole   1$^{\circ}$10$\arcmin$39$\arcsec$\colh  \coli  \colj  
\colk  \coll  \colm  \coln   26\colo   4\colp    4.7\colq   2.5\colr \eol
\cola  65\ \colc  29$^m$ 4.6\cole   1$^{\circ}$20$\arcmin$02$\arcsec$\colh  \coli  \colj  
\colk  \coll  \colm  \coln   33\colo   3\colp    8\colq   2\colr \eol
\cola  66\ \colc  29$^m$ 4.7\cole   1$^{\circ}$16$\arcmin$15$\arcsec$\colh  \coli  \colj  
\colk  \coll  \colm  \coln   56\colo   6\colp   15\colq   2\colr \eol
\cola  69\ \colc  29$^m$ 5.0\cole   1$^{\circ}$22$\arcmin$04$\arcsec$\colh  \coli  \colj  
\colk  \coll  \colm  \coln   28\colo   3\colp    7\colq   2\colr \eol
\cola  74\ \colc  29$^m$ 8.1\cole   1$^{\circ}$12$\arcmin$35$\arcsec$\colh  \coli  \colj  
\colk  \coll  \colm  \coln   13\colo   2\colp    2.4\colq   1.2\colr \eol
\cola  80\ \colc  29$^m$ 9.6\cole   1$^{\circ}$09$\arcmin$29$\arcsec$\colh  \coli  \colj  
\colk  \coll  \colm  \coln   12\colo  18\colp    4.1$^d$\colq   2.0\colr \eol
\cola  82\ \colc  29$^m$10.6\cole   1$^{\circ}$18$\arcmin$30$\arcsec$\colh  \coli  \colj  
\colk  \coll  \colm  \coln   27\colo   3\colp    6.3\colq   1.8\colr \eol
\cola  84\ \colc  29$^m$10.6\cole   1$^{\circ}$18$\arcmin$06$\arcsec$\colh  \coli  \colj  
\colk  \coll  \colm  \coln   12\colo   2\colp    3.8\colq   2.5\colr \eol
\cola  91\ \colc  29$^m$12.4\cole   1$^{\circ}$11$\arcmin$20$\arcsec$\colh  \coli  \colj  
\colk  \coll  \colm  \coln   23\colo   3\colp    5.3\colq   1.9\colr \eol
\cola  94\ \colc  29$^m$13.4\cole   1$^{\circ}$11$\arcmin$59$\arcsec$\colh  \coli  \colj  
\colk  \coll  \colm  \coln  118\colo  12\colp   33\colq   4\colr \eol
\cola  96\ \colc  29$^m$14.2\cole   1$^{\circ}$21$\arcmin$30$\arcsec$\colh  \coli  \colj  
\colk  \coll  \colm  \coln   31\colo   4\colp    8.3\colq   2.3\colr \eol
\cola  98\ \colc  29$^m$14.9\cole   1$^{\circ}$21$\arcmin$55$\arcsec$\colh  \coli  \colj  
\colk  \coll  \colm  \coln   79\colo  13\colp   19\colq   4\colr \eol
\cola 102\ \colc  29$^m$15.8\cole   1$^{\circ}$ 9$\arcmin$32$\arcsec$\colh  \coli  \colj  
\colk  \coll  \colm  \coln  128\colo  22\colp   34\colq   5\colr \eol
\cola 104\ \colc  29$^m$16.9\cole   1$^{\circ}$18$\arcmin$37$\arcsec$\colh  \coli  \colj  
\colk  \coll  \colm  \coln   28\colo   2\colp    7.0\colq   2.0\colr \eol
\cola 112\ \colc  29$^m$19.9\cole   1$^{\circ}$20$\arcmin$59$\arcsec$\colh  \coli  \colj  
\colk  \coll  \colm  \coln   65\colo   8\colp   38\colq   5\colr \eol
\cola 126\ \colc  29$^m$25.6\cole   1$^{\circ}$03$\arcmin$50$\arcsec$\colh  \coli  \colj  
\colk  \coll  \colm  \coln   59\colo   2\colp   12\colq   4\colr \eol
\cola 132\ \colc  29$^m$26.8\cole   1$^{\circ}$11$\arcmin$56$\arcsec$\colh  \coli  \colj  
\colk  \coll  \colm  \coln   21\colo   2\colp    5.0\colq   2.7\colr \eol
\cola 134\ \colc  29$^m$27.3\cole   1$^{\circ}$12$\arcmin$58$\arcsec$\colh  \coli  \colj  
\colk  \coll  \colm  \coln   15\colo   2\colp    1.4\colq   2.5\colr BD+01 3686\eol
\cola 140\ \colc  29$^m$28.6\cole   1$^{\circ}$10$\arcmin$24$\arcsec$\colh  \coli  \colj  
\colk  \coll  \colm  \coln   51\colo   2\colp   15\colq   2\colr \eol
\cola 144\ \colc  29$^m$29.4\cole   1$^{\circ}$14$\arcmin$03$\arcsec$\colh  \coli  \colj  
\colk  \coll  \colm  \coln   49\colo   7\colp   12\colq   2\colr \eol
\cola 145\ \colc  29$^m$29.7\cole   1$^{\circ}$13$\arcmin$09$\arcsec$\colh  \coli  \colj  
\colk  \coll  \colm  \coln   37\colo   6\colp    9.0\colq   2.5\colr \eol
\cola 153\ \colc  29$^m$30.9\cole   1$^{\circ}$15$\arcmin$18$\arcsec$\colh  \coli  \colj  
\colk  \coll  \colm  \coln   10\colo   2\colp    0.9\colq   2.1\colr \eol
\cola 166\ \colc  29$^m$32.6\cole   1$^{\circ}$09$\arcmin$46$\arcsec$\colh  \coli  \colj  
\colk  \coll  \colm  \coln   20\colo   2\colp    7.0\colq   2.6\colr \eol
\cola 184\ \colc  29$^m$37.2\cole   1$^{\circ}$09$\arcmin$15$\arcsec$\colh  \coli  \colj  
\colk  \coll  \colm  \coln  125\colo   3\colp   37\colq   6\colr \eol
\cola 189\ \colc  29$^m$37.7\cole   1$^{\circ}$11$\arcmin$18$\arcsec$\colh  \coli  \colj  
\colk  \coll  \colm  \coln   67\colo   2\colp   31\colq   2\colr \eol
\cola 190\ \colc  29$^m$37.6\cole   1$^{\circ}$11$\arcmin$30$\arcsec$\colh  \coli  \colj  
\colk  \coll  \colm  \coln   68\colo   10\colp   32\colq   3\colr \eol
\cola 191\ \colc  29$^m$37.9\cole   1$^{\circ}$00$\arcmin$36$\arcsec$\colh  \coli  \colj  
\colk  \coll  \colm  \coln   44$^c$\colo   2\colp   10\colq   3\colr \eol
\cola 208\ \colc  29$^m$41.0\cole   1$^{\circ}$12$\arcmin$40$\arcsec$\colh  \coli  \colj  
\colk  \coll  \colm  \coln   36\colo   6\colp    9\colq   4\colr SVS18\eol
\cola 209\ \colc  29$^m$41.1\cole   1$^{\circ}$20$\arcmin$56$\arcsec$\colh  \coli  \colj  
\colk  \coll  8.69\colm $<$0.01\coln   44\colo   5\colp   12\colq   2\colr \eol
\cola 210\ \colc  29$^m$41.5\cole   1$^{\circ}$10$\arcmin$00$\arcsec$\colh  \coli  \colj  
\colk  \coll  \colm  \coln   12\colo   2\colp    6\colq   4\colr \eol
\cola 228\ \colc  29$^m$45.2\cole   1$^{\circ}$18$\arcmin$46$\arcsec$\colh  \coli  
\colj 10.09 \colk  \coll  8.67\colm $<$0.01\coln   91\colo   9\colp   19\colq   
3\colr SVS 9\eol
\cola 238\ \colc  29$^m$47.6\cole   1$^{\circ}$05$\arcmin$12$\arcsec$\colh  \coli  
\colj  \colk  \coll  \colm  \coln 1414\colo  19\colp  285\colq   7\colr StRS 208\eol
\cola 243\ \colc  29$^m$48.9\cole   1$^{\circ}$11$\arcmin$43$\arcsec$\colh 
11.18\coli $<$0.01\colj  9.53 \colk $<$0.01\coll  8.84\colm $<$0.01\coln   
35\colo   4\colp    8\colq   3\colr SVS3\eol
\cola 244\ \colc  29$^m$49.2\cole   0$^{\circ}$58$\arcmin$54$\arcsec$\colh  \coli  
\colj  \colk  \coll  \colm  \coln   20\colo   3\colp    7$^e$\colq   3\colr \eol
\cola 255\ \colc  29$^m$49.9\cole   1$^{\circ}$09$\arcmin$24$\arcsec$\colh  \coli  
\colj  \colk  \coll  \colm  \coln   91\colo   2\colp   24\colq   4\colr \eol
\cola 264\ \colc  29$^m$51.6\cole   0$^{\circ}$56$\arcmin$21$\arcsec$\colh  \coli  
\colj  \colk  \coll  \colm  \coln   14\colo   2\colp    5\colq   1\colr \eol
\cola 267\ \colc  29$^m$51.6\cole   1$^{\circ}$10$\arcmin$32$\arcsec$\colh  \coli  
\colj  \colk  \coll  \colm  \coln   48\colo   3\colp   13\colq   6\colr SVS 5\eol
\cola 271\ \colc  29$^m$52.7\cole   1$^{\circ}$04$\arcmin$38$\arcsec$\colh  \coli  
\colj  \colk  \coll  \colm  \coln   24\colo   6\colp    4\colq   6\colr \eol
\cola 290\ \colc  29$^m$54.3\cole   1$^{\circ}$17$\arcmin$47$\arcsec$\colh  \coli  
\colj  \colk  \coll  \colm  \coln   27\colo   5\colp    4\colq   2\colr \eol
\cola 296\ \colc  29$^m$55.8\cole   1$^{\circ}$04$\arcmin$14$\arcsec$\colh  \coli  
\colj  \colk  \coll  \colm  \coln  118\colo   5\colp   31\colq   4\colr \eol
\cola 300\ \colc  29$^m$56.1\cole   1$^{\circ}$00$\arcmin$24$\arcsec$\colh  \coli  
\colj  \colk  \coll  \colm  \coln   53\colo  19\colp   12\colq  13\colr \eol
\cola 311\ \colc  29$^m$57.6\cole   1$^{\circ}$10$\arcmin$46$\arcsec$\colh  \coli  
\colj  \colk  \coll  \colm  \coln   26\colo   2\colp   15\colq   3\colr 
BD+01 3689B\eol
   \noalign{\smallskip}
   \hline  
   \end{tabular}
\]

         {\small (continued on next page)} \\
         
   \end{table*}
   
   \begin{table*}

\newcommand\cola {\null}
\newcommand\colc {&}
\newcommand\cole {&}
\newcommand\colh {&}
\newcommand\coli {&}
\newcommand\colj {&}
\newcommand\colk {&}
\newcommand\coll {&}
\newcommand\colm {&}
\newcommand\coln {&}
\newcommand\colo {&}
\newcommand\colp {&}
\newcommand\colq {&}
\newcommand\colr {&}
\newcommand\eol{\\}
\newcommand\extline{&&&&&&&&&&&&&\eol}

\[
  \begin{tabular}{rccrrrrrrrrrrl}
  \hline
  \noalign{\smallskip} 

{ISO} & $\alpha_{2000}$ & $\delta_{2000}$ & {$J$} & {$\sigma_{J}$} & {$H$} 
& $\sigma_{H}$ & {$K$} & $\sigma_{K}$ & $F_{\nu}^{6.7}$ & $\sigma_{6.7}$ 
& $F_{\nu}^{14.3}$ & $\sigma_{14.3}$ & {Other ID} \\
\# & 18$^h$ &  & mag & & mag & & mag &  & mJy & mJy & mJy & mJy & \\
\hline
\cola 324 \colc  29$^m$58.4\cole   1$^{\circ}$20$\arcmin$26$\arcsec$\colh  \coli  
\colj  \colk  \coll  \colm  \coln   15\colo   3\colp    4\colq   3\colr SVS 16\eol
\cola 332 \colc  30$^m$ 0.2\cole   1$^{\circ}$09$\arcmin$47$\arcsec$\colh  \coli  
\colj  \colk  \coll  \colm  \coln   62\colo   2\colp   36\colq   4\colr \eol
\cola 334 \colc  30$^m$ 0.0\cole   1$^{\circ}$21$\arcmin$55$\arcsec$\colh  \coli  
\colj  \colk  \coll  \colm  \coln  160\colo   10\colp   40\colq   3\colr SVS 15\eol
\cola 337 \colc  30$^m$ 0.6\cole   1$^{\circ}$15$\arcmin$19$\arcsec$\colh 19.13\coli 
0.279\colj 12.61 \colk 0.004\coll  8.92\colm 0.001\coln  340\colo   8\colp   
95\colq   6\colr CK2,EC118\eol
\cola 339 \colc  30$^m$ 0.8\cole   1$^{\circ}$19$\arcmin$41$\arcsec$\colh  \coli  
\colj  \colk  \coll  \colm  \coln   70\colo   7\colp   15\colq   2\colr SVS 13\eol
\cola 342 \colc  30$^m$ 1.6\cole   0$^{\circ}$59$\arcmin$30$\arcsec$\colh  \coli  
\colj  \colk  \coll  \colm  \coln   40\colo   9\colp    8\colq   6\colr \eol
\cola 349 \colc  30$^m$ 3.7\cole   1$^{\circ}$ 4$\arcmin$52$\arcsec$\colh  \coli  
\colj  \colk  \coll  \colm  \coln   97\colo   6\colp   21\colq   9\colr \eol
\cola 372 \colc  30$^m$ 8.4\cole   0$^{\circ}$55$\arcmin$29$\arcsec$\colh  \coli  
\colj  \colk  \coll  \colm  \coln   80\colo   2\colp   19\colq   4\colr \eol
\cola 375 \colc  30$^m$ 8.7\cole   0$^{\circ}$55$\arcmin$16$\arcsec$\colh  \coli  
\colj  \colk  \coll  \colm  \coln   98\colo   3\colp   28\colq   5\colr \eol
\cola 378 \colc  30$^m$ 9.0\cole   1$^{\circ}$14$\arcmin$44$\arcsec$\colh 11.81\coli 
0.003\colj  9.68 \colk 0.001\coll  8.67\colm 0.001\coln   70\colo   7\colp   
14\colq   4\colr SVS7\eol
\cola 382 \colc  30$^m$ 9.2\cole   1$^{\circ}$17$\arcmin$53$\arcsec$\colh  \coli  
\colj  \colk  \coll  \colm  \coln    7\colo   2\colp    3\colq   4\colr \eol
\cola 389 \colc  30$^m$10.4\cole   1$^{\circ}$19$\arcmin$36$\arcsec$\colh  \coli  
\colj  \colk  \coll  \colm  \coln   18\colo   2\colp    5\colq   4\colr BD+01 3693\eol
\cola 401 \colc  30$^m$12.1\cole   1$^{\circ}$16$\arcmin$41$\arcsec$\colh 11.39\coli 
0.003\colj  9.59 \colk 0.001\coll  8.64\colm 0.001\coln   66\colo  11\colp   
13\colq   5\colr SVS6\eol
\cola 409 \colc  30$^m$14.7\cole   1$^{\circ}$05$\arcmin$24$\arcsec$\colh  \coli  
\colj  \colk  \coll  \colm  \coln   15\colo   3\colp    5\colq   3\colr \eol
\cola 410 \colc  30$^m$14.7\cole   1$^{\circ}$06$\arcmin$17$\arcsec$\colh  \coli  
\colj  \colk  \coll  \colm  \coln   22\colo   2\colp    7\colq   3\colr \eol
   \noalign{\smallskip}
   \hline
   \end{tabular}
\]

$^{b}$ Binary or close companion not quite resolved by ISOCAM. \\
$^{c}$ Flux measurement affected by proximity to the dead column. If 
       the dead column cannot be avoided by any of the redundant 
       observations, this flagging. If source is located on the dead 
       column, a flux measurement is not attempted at all. \\
$^{d}$ Source close to the detector edge in all redundant observations. \\
$^{e}$ Extended source. \\
        
    Identifier acronyms are related to the following references: 
    SVS (Strom et al. 1976); CK (Churchwell \& Koornneef 1986); 
    StRS (Stephenson 1992); EC (Eiroa \& Casali 1992), SCB (Straizys et 
    al. 1996).
        
   \end{table*}

Except for a few transition objects, the ``blue'' group (triangles) 
is mainly located at the colour index of normal photospheres, i.e. 
$[14.3/6.7] = -0.66$ or $\alpha_{\rm IR}^{7-14} = -3$ (dashed line). The 
spread around this value indicates the increasing photometric uncertainty 
with decreasing flux, although some of the scatter might be real. M giants 
have intrinsic excesses in the colour index $[14.3/6.7]$ of less than 0.1, 
while late M giants are expected to have excesses of up to $\sim 0.2$ above 
normal photospheres. Although late M giants are few, a number of M giants 
are expected in the observed sample of field stars. Thus, the slight 
displacement of this group of sources above the colour of normal 
photospheres could be due to a combination of extinction and intrinsic 
colours of M giants. The effect of extinction is small, however. A reddening 
vector of size corresponding to A$_K = 3$ is indicated in the figure. 
(More about extinction in Sect.~\ref{nir}.)

This same dichotomy in colour was found also in the Chamaeleon dark clouds
\citep{nor96}, in RCrA \citep{olo99}, and in $\rho$ Ophiuchi \citep{bon01}. 
According to the SED index, $\alpha_{\rm IR}^{7-14}$, the mid-IR excess 
sources are Class~II and Class~I types of Young Stellar Objects (YSOs). 
In Chamaeleon~I about 40\% of the mid-IR excess sources had been previously 
classified as Classical T Tauri stars (CTTS), and in $\rho$ Ophiuchi $\sim$ 
50\% were previously known as Class~II and Class~Is. In Serpens, very few 
sources are optically visible and the IRAS data suffer badly from source 
confusion. The known YSOs are thus a sample of sources classified from 
near-IR excesses,
association with nebulosity, emission-line stars, ice features, clustering
properties, and variability \citep{eir92,hod96,hor97,sog97,gio98,kaa99a}.
Previously known YSOs are not found in the ``blue'' group, although 
Class\,III sources, young stars with marginal or no IR excess, have their 
locus there. A few objects in the transition phase between Class~II and
Class~III are evident in the ``blue'' group. Class\,III sources without 
IR excess cannot be distinguished from field stars purely on the basis of 
broad band infrared photometry. They can, however, be efficiently identified 
through X-ray observations. 

The fact that intrinsic IR excess is so easily distinguished from 
reddening at these wavelengths, enables an unambiguous classification of 
the youngest YSOs on the basis of one single colour index $[14.3/6.7]$. 
In this way ISOCAM found $53$ Class~I and Class~II YSOs, of which only 
$28$ had been previously suggested as YSO candidates.

Positions and photometry of the 53 Class~I and Class~II YSOs are listed 
among other members of the Serpens Core cluster in Table~\ref{tbl-2}. 
The 71 ``blue'' sources are listed in Table~\ref{tbl-3}. Many of these are 
field stars, but some are likely to be young cluster members of Class~III 
type. Fig.~\ref{fig1} shows that the CW region contains as many as 20 (10) 
sources detected at 6.7 $\mu$m (14.3 $\mu$m) above the expected field star 
contribution according to the galactic model by \citet{wai92}. Because there 
are only two IR excess sources in the CW field, of which one is very faint, these 
are practically all belonging to the ``blue'' sample.
Since the total region surveyed is about 3 times as large as the CW field, 
perhaps more than 20 ``blue'' sources from Table~\ref{tbl-3} could be 
Class\,IIIs belonging to the cluster.


\section{Near-IR and mid-IR excess YSOs}
\label{nir}

\begin{figure}
\resizebox{\hsize}{!}{\includegraphics{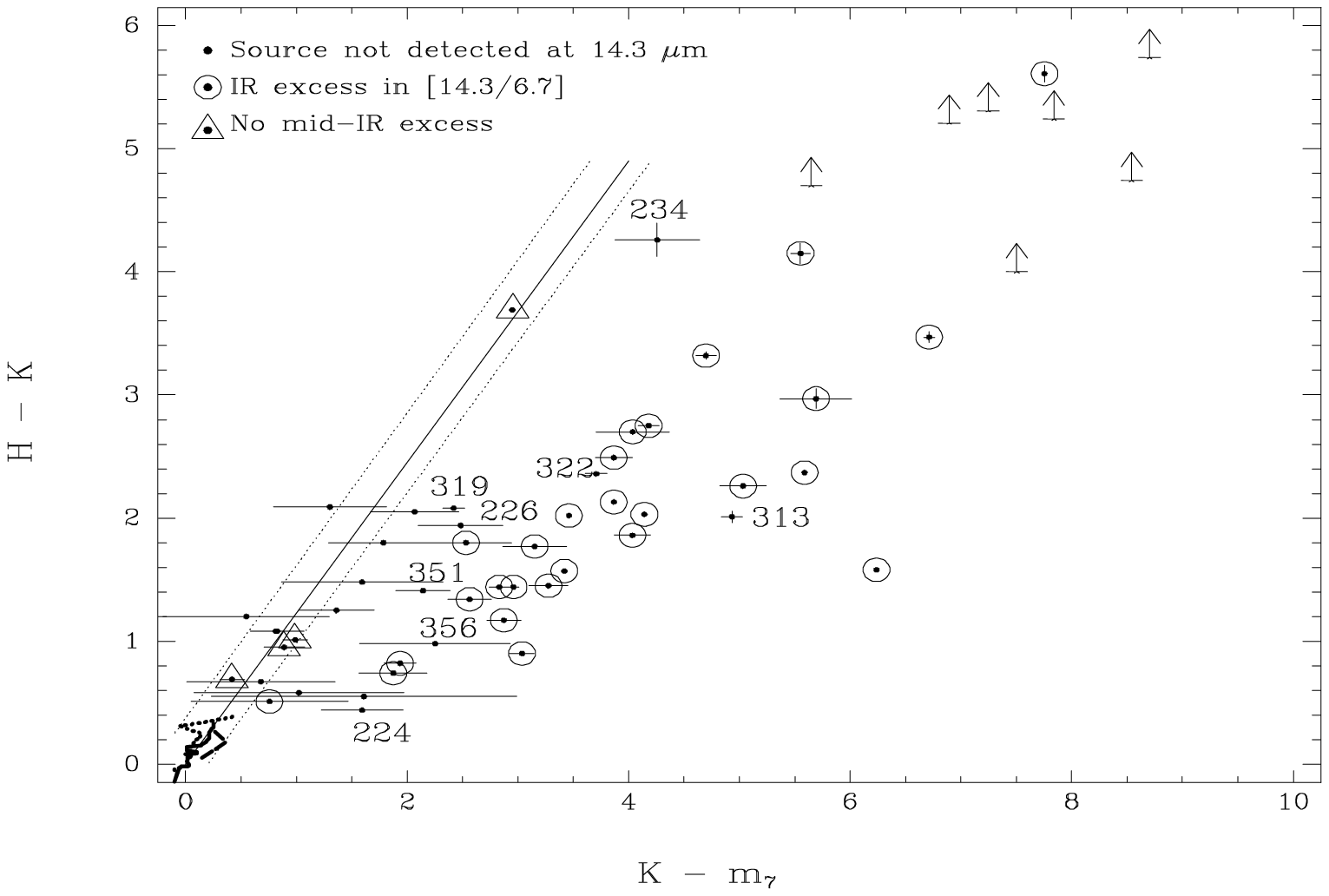}}
\caption{The $H-K/K-m_{\rm 6.7}$ diagram of the Serpens Cloud Core with 
         individual error bars for each source, and some $H$ upper limits 
         (arrows). The IR excess sources from Fig.~\ref{fig-323} are 
         encircled. The intrinsic colours of main-sequence (bold solid), 
         giants (bold dotted), and supergiant (bold dashed) stars cluster 
         close to (0,0). A reddening vector (solid line) is estimated on 
         the basis of the four sources without $[14.3/6.7]$ excess 
         (triangles). Eight additional IR excess YSOs (numbered) were found 
         in this diagram, i.e. displaced by more than 1$\sigma$ to the right
         of the reddening band (dotted vector).
         \label{fig-hk2}}
\end{figure}

When it comes to detecting IR excesses, the $J$ band usually limits the number 
of sources in a $J-H/H-K$ diagram because of its sensitivity to extinction. 
For the ISOCAM observations the 14.3 $\mu$m band is the less sensitive, unless 
the sources are extremely red (see limits in Fig.~\ref{fig-323}). Using the
$H-K/K-m_{6.7}$ diagram as in \citet{olo99}, however, we should be able to
detect more IR excess objects. Fig.~\ref{fig-hk2} shows that practically all 
sources with excesses in the colour index $[14.3/6.7]$ are also found to have 
excesses in their $K - m_{6.7}$ index. Approximate intrinsic colours of 
main-sequence, giant and supergiant stars, taken from the source table of 
\citet{wai92} and interpolated between 2.2 and 12 $\mu$m, are given as boldface 
curves (solid, dotted and dashed, respectively). 

Sources with ISOCAM mid-IR excesses (circles) separate well from those 
without mid-IR excesses (triangles). A reddening vector has been calculated 
by fitting a line from origin through the 4 sources without mid-IR excesses. 
The slope (1.23) is mainly constrained by the star CK2, which is believed 
to be a background supergiant, see e.g. \citet{cas96}. On the basis of a 
number of background stars in the $J-H/H-K$ diagram, \citet{kaa99a} found a
reddening law of the form A$_{\lambda} \sim \lambda^{-1.9}$ to fit the Serpens
data. Extrapolation of this law to 6.7 $\mu$m would give a slope of 0.83 in 
the $H-K/K-m_{\rm 6.7}$ diagram, which is in disagreement with all the 4
sources without mid-IR excess. The A$_{\lambda} \sim \lambda^{-1.7}$ law
\citep{whi88} gives an even shallower slope. By assuming, however, the
A$_{\lambda} \sim \lambda^{-1.9}$ law to hold for $H$ and $K$, and using 
the empirical slope found in Fig.~\ref{fig-hk2}, the extinction in the 6.7
$\mu$m band is estimated to be A$_{\rm 6.7} = 0.41 $ A$_{K}$. 
This is in good agreement with the values 0.37, 0.35 and 0.47 found by \citet{jia03}
from the ISOGAL survey, whose three values depend on the actual form of 
the near-IR extinction curve applied.


As shown in \citet{kaa00} only about 50\% of the IR excess sources which 
display clear excesses in the single ISOCAM colour index $[14.3/6.7]$ show 
up as IR excess objects in the $J-H/H-K$ diagram. This was demonstrated 
by plotting the same sources in the two diagrams, using a statistically 
significant sample of several hundred sources in Serpens, $\rho$ Ophiuchi, 
Chamaeleon~I, and RCrA. This result implies that a sole use of the $J-H/H-K$ 
diagram will severely underestimate the IR-excess population of YSOs in active
star formation regions. \citet{bon01} showed that in $\rho$ Ophiuchi only
50\% of the 123 Class II sources have near-IR excesses large enough to be
recognizable in the $J-H/H-K$ diagram. Similarly we confirm here specifically
for Serpens that the mid-IR excess sources cluster along the reddening line
when plotted in a $J-H/H-K$ diagram, and that only 50\% would be recognized 
as having IR excesses from JHK data alone.

The $H-K/K-m_{7}$ diagram is more efficient than the $J-H/H-K$ diagram in 
distinguishing between intrinsic circumstellar excess (circles) and reddening 
(triangles). Thus, by sampling the SED a bit more into the mid-IR, the use 
of two colour indices to select sources with intrinsic IR excesses becomes 
substantially less prone to the influence from cloud extinction. 
Fig.~\ref{fig-hk2} also shows that the mid-IR excesses apparent from the 
$[14.3/6.7]$ index, always appear already at 6.7 $\mu$m, but only in half 
of the cases at 2.2 $\mu$m.

In addition to the $53$ IR excess sources obtained from Fig~\ref{fig-323}, 
another eight sources were found to have IR excess from the $H-K/K-m_{7}$ 
diagram, i.e. displaced by more than 1$\sigma$ to the right of the reddening 
band (dotted line).  This means we find 25\% more IR excess sources within 
the centre region than by using the $[14.3/6.7]$ index alone. 

Class\,II sources are most likely Classical T Tauri stars (CTTS), but we can
not exclude that some of them are weak-lined T Tauri stars (WTTS). Among
the Class\,II sources detected with ISOCAM in Chamaeleon\,I about 1/3 had been
previously classified as WTTS \citep{nor96}. 
For WTTS one could attribute the presence of mid-IR excess but a lack of 
near-IR excess to an inner hole in the circumstellar disk \citep{mon99}.  
CTTS, on the other hand, are believed to have inner disks, since they show 
strong H$\alpha$ emission, which is commonly interpreted as a signature of 
the accretion process onto the surface of the object, but could also arise
in stellar winds. Optical information is scarce in the Serpens Cloud Core 
because of the large cloud extinction, and we do not know what fraction of 
the Class\,IIs are strong H$\alpha$ emitters. 
It is well known from studies of CTTS locations in the $J-H/H-K$ diagram 
\citep{lad92,mey97}, that many CTTS lack detectable near-IR excesses; about 
40\% in Taurus-Auriga according to \citet{str93}. 

For typical CTTS the near-IR wavelength region is strongly dominated by the 
photospheric emission, and rather large amounts of dust hot enough to produce 
a strong excess at 2 $\mu$m are therefore needed in order to distinguish 
intrinsic IR excess from the effects of scattering and extinction in the 
$J-H/H-K$ diagram. In a region such as the Serpens Cloud Core, where 7 of 
the Class\,IIs in question here were originally proposed to be YSOs due to 
their association with nebulosity \citep{eir92}, it is likely that
scattered light in the J and H bands adds to the lack of detectable
near-IR excesses as it gives a bluer $H-K$ colour index. 

From an ISOCAM sample of CTTS with mid-IR excesses in Chamaeleon\,I, 
\citet{com00} found evidence for the presence of near-IR excesses to be 
correlated with luminosity, suggesting an incapability of objects with very 
low temperatures and luminosities (young brown dwarfs and very low mass CTTS) 
to raise the temperature needed at the inner part of the circumstellar disk 
to produce a detectable excess at 2.2 $\mu$m, in agreement with model 
predictions \citep{mey97}. 
The majority of the Serpens sources are substantially more luminous. As pointed 
out by \citet{hil98}, the larger the stellar radius (i.e. the younger the star 
is), the more difficult it is to separate near-IR excesses from the stellar 
flux. This could perhaps be part of the explanation for the youngest sources 
(e.g. two flat-spectrum sources without near-IR excess). But we found all the
Class\,I sources to possess near-IR excesses, such that the role of larger 
radii in these cases seems to be well compensated for, probably by their 
larger disk accretion rates. 
A statistical interpretation based on broad band photometry is likely 
over-simplified, however, and there are many properties intrinsic to the 
star-disk system (such as disk inclination angle, disk accretion rate, 
stellar mass and radius) which contribute to the complexity of the 
individual YSOs. From the results presented here for Serpens, in 
\citet{bon01} for $\rho$ Ophiuchi, and in \citet{kaa00} for the ISOCAM star 
formation surveys in general, it is evident that one should be careful in 
estimating disk frequencies among YSOs based on near-IR excesses only, 
see also \citet{hai01}. Since our study samples only IR excess objects, we 
cannot say anything about the disk fraction among YSOs in Serpens.


\begin{figure}
\resizebox{\hsize}{!}{\includegraphics{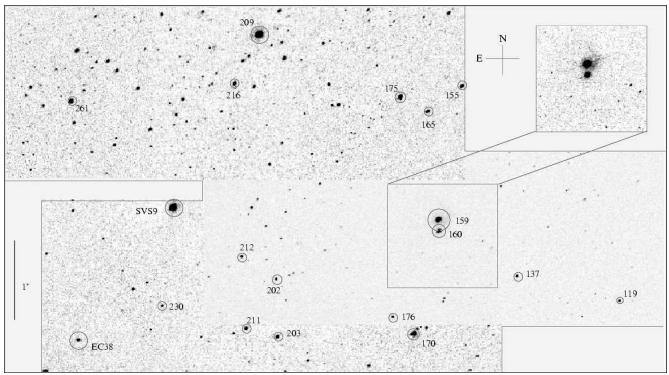}}
\caption{Arnica $K$ band mapping of the region to the NW of the $JHK$ field. 
         ISOCAM sources are encircled and their identifications given. 
         Sources 159, 202, 216 and EC38 are YSOs with $[14.3/6.7]$ 
         colour excess. 
         Sources SVS9 and 209 are without such an excess. The remaining sources 
         are not detected at 14.3 $\mu$m. See Sect.~\ref{ind}.
         \label{fig-nw}}
\end{figure}


\section{Characterization of the YSO population}
\label{char}

\subsection{Large fraction of Class\,I sources}
\label{classI}

\begin{figure}
\resizebox{\hsize}{!}{\includegraphics{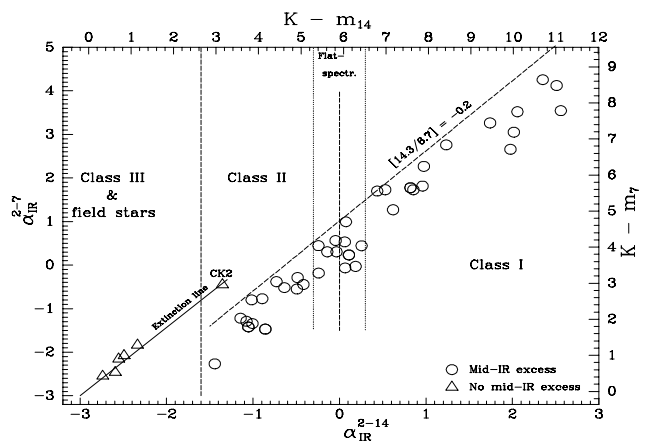}}
\caption{The spectral index $\alpha_{\rm IR}^{2-7}$ versus 
         $\alpha_{\rm IR}^{2-14}$, or equivalently, the $K-m_{7}/K-m_{14}$
         colour diagram, for a subset of the ISOCAM sample. 
         The locations of Class I, Class II and Class III sources and 
         field stars are indicated. Note the relatively large fraction 
         of Class I sources in this diagram.\label{fig-k2k3}}
\end{figure}

We have combined the photometry from ISOCAM with the available $K$ band 
photometry from 1996\footnote{This means the central 8$\arcmin \times$ 
6$\arcmin$ JHK field and the NW field shown in Fig.~\ref{fig-nw}.}
\citep{kaa99a} and calculated the two SED indices: $\alpha_{\rm IR}^{2-14}$ 
and $\alpha_{\rm IR}^{2-7}$, which are plotted against each other in 
Fig.~\ref{fig-k2k3} for the 39 mid-IR excess sources and the 6 sources 
without mid-IR excesses. The index $\alpha_{\rm IR}^{2-14}$ is close to 
the index originally used to define the three classes I, II and III from 
the shape of their SED between 2.2 and 10 (or 25) $\mu$m by 
\citet{lad84,lad87}. 
Figure~\ref{fig-k2k3} indicates the loci of these classes in addition to 
a transitional class referred to as "flat spectrum" sources. 

According to the updated IR spectral classification scheme \citep{and94,gre94}
we tentatively define YSOs with $\alpha_{\rm IR}^{2-14} > 0.3$ as Class\,I 
sources, those with $-0.3 < \alpha_{\rm IR}^{2-14} < 0.3$ as flat-spectrum 
sources, and objects with $-1.6 < \alpha_{\rm IR}^{2-14} < -0.3$ as Class\,II 
sources. 
There is a marginal hint of a gap in the distribution of sources along
the $[14.3/6.7] = -0.2$ axis at $\alpha_{\rm IR}^{2-14} \approx  0.3$, which 
was seen also in the $\rho$ Ophiuchi sample \citep{bon01}, though at $\approx 
0.5$. 
It is apparent that we cannot distinguish between Class III sources and 
field stars. (The transition sources in the ``blue'' group (see 
Fig.~\ref{fig-323}) with some mid-IR excesses are all located outside the 
field for which we have K-band observations.) 

IR excess sources without detection at 14.3 $\mu$m (see Sec.~4.2) can be 
classified from the approximately linear relation between 
$\alpha_{\rm IR}^{2-14}$ and $\alpha_{\rm IR}^{2-7}$ found in 
Fig.~\ref{fig-k2k3}. In this case Class\,Is are those which have 
$\alpha_{\rm IR}^{2-7} > 1.2$, corresponding to $K-m_{7} > 4.8$, Class\,IIs 
are those which have $\alpha_{\rm IR}^{2-7} < -0.25$, corresponding to 
$K-m_{7} < 3.2$, and flat-spectrum sources are those in between. Thus, the
total number of sources in each category is larger than apparent from 
Fig.~\ref{fig-k2k3}. From Fig.~\ref{fig-hk2} we have 1 more Class~I, 2 more 
flat-spectrum sources, and 5 more Class~IIs. Three sources (260, 277 and 308)
have only $K$ and 6.7 $\mu$m detections, but their very red colours strongly 
suggest membership in the Class~I category. 


The number of Class\,I sources ($19$) is thus about equal to the number of 
Class\,II sources ($18$) in the central part of the Serpens cluster. This is 
very unusual since Class\,II sources 
outnumber Class\,I sources by typically 10 to 1 in star formation regions. 
If we include the $13$ "flat spectrum" sources in the Class\,II group, the
Class\,I/Class\,II number ratio is still as high as $19/31$. This is 
exceptional, also compared to the results obtained from ISOCAM surveys in 
other regions; in Chamaeleon I this number ratio is $5/{\bf 42}$ \citep{kaa99b} 
and in $\rho$ Ophiuchi: $16/123$ \citep{bon01}. Such a large population of 
Class\,I sources indicates a recent burst of star formation in this region 
and would be in line with the rich collection of Class\,0 objects found in 
this cluster by \citet{cas93} and \citet{hur96}. 

\subsection{Reddening effect on the $\alpha_{\rm IR}^{2-14}$ index?}
\label{redden}

While a spectral index between two fixed wavelengths, such as 2.2 and 14.3 
$\mu$m, is practical and provides a preliminary classification for a large 
number of sources, a truly reddening-independent classification is obtainable 
only when complete SEDs are available over the $\sim$ 2 to 100 $\mu$m range 
(or to 1.3 mm if Class\,0 sources are involved). In the following we discuss 
the possibility that sources defined as Class\,Is from the 
$\alpha_{\rm IR}^{2-14}$ index could be heavily extinguished Class\,II or 
flat-spectrum sources.

The source CK2 (probably a background supergiant) is seen through more than 
50 magnitudes of visual extinction \citep{kaa99a}, and indicates 
an empirical slope (=1.6) of the extinction law in Fig.~\ref{fig-k2k3}, 
as do five other sources without mid-IR excesses. This reddening 
line translates to a slope of $0.92$ in the $K-m_{7}/K-m_{14}$ diagram, 
which gives the relation A$_{\rm 14}$ = 0.88 A$_{\rm 7}$ 
in terms of magnitudes\footnote{The same was found in Cha I (unpublished) 
as well as in $\rho$ Ophiuchi \citep{bon01}. This implies that, for all 
these star formation regions, the extinction is slightly larger at 6.7 
$\mu$m than at 14.3 $\mu$m, see also \citet{olo99}. Hence, 
we see no minimum in the extinction law in the 4-8 $\mu$m interval, as 
expected for standard graphite-silicate mixtures \citep{mat77,dra84}. 
But our results agree with the extinction law found towards the Galactic
centre by \citet{lut96} and the ISOGAL results in \citet{jia03} who find
A$_{\rm 14}$ = 0.87 A$_{\rm 7}$ when using the Rieke \& Lebofsky law
updates\citep{rie89} for the near-IR extinction.}, 
by using the relation $A_{7} = 0.41 A_K$ obtained in Sect.~\ref{nir}. 
It is remarkable that the group of mid-IR excess sources are lined up along 
the same slope as the one outlined by the reddened field stars. In this 
sense, some of the sources classified as Class\,I objects on the basis of
Fig.~\ref{fig-k2k3} could be interpreted as highly extinguished Class\,II 
objects, either deeply embedded in the cloud or seen edge-on. 
Flat-spectrum sources and Class\,IIs are generally expected to be optically 
visible objects. In Serpens, however, only one (ISO-Ser-159) of the 13 
flat-spectrum sources and only three of the 18 Class\,IIs are optically 
visible. This shows the extreme degree of embeddedness, and is in agreement 
with the C$^{18}$O map of \citet{whi95}, which suggests that the visual cloud 
extinction is larger than 30 magnitudes everywhere in the NW-SE ridge and may 
reach values as high as $A_V \sim 200$ in some places.
The fact that the slopes are so similar suggests that Class\,Is are either 
merely more embedded Class\,IIs, or that the extinction behaves roughly in 
the same way for Class\,I envelopes as for the cloud in general at these 
wavelengths. This would be in line with \citet{pad99} who observed three 
Class\,I SED YSOs with HST/NICMOS and found edge-on disks in all three cases. 

\begin{figure}
\resizebox{\hsize}{!}{\includegraphics{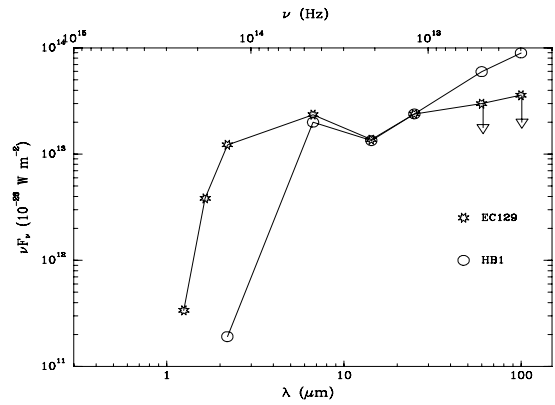}}
\caption{SEDs of HB1 and EC129. The flux points at 25, 60 and 100 $\mu$m
are taken from Hurt \& Barsony (1996). Note that the 60 and 100 $\mu$m 
points for EC129 are upper limits only.  \label{fig-sed}}
\end{figure}

As shown in Fig.~\ref{fig-323} the $\alpha_{\rm IR}^{7-14}$ index is rather 
insensitive to extinction. 
As many as 15 YSOs in our sample have $\alpha_{\rm IR}^{2-14} > 0$, but
$\alpha_{\rm IR}^{7-14} < 0$, among them the well known DEOS. For some of 
these we have more spectral information than the three flux points at 2.2, 
6.7 and 14.3 $\mu$m. The two sources HB1 and EC129 are relatively isolated 
YSOs both in the near-IR and mid-IR images and should not suffer so much 
from the source confusion that sets a limit to the usefulness of IRAS data 
for the other sources. We use HIRES IRAS fluxes and upper limits from
\citet{hur96} and extend our mid-IR SEDs to longer wavelengths.
Figure~\ref{fig-sed} shows that both HB1 and EC129 (HB2) have rising spectra 
beyond a clear dip in the SED at $\sim$ 14 $\mu$m. These sources have clearly 
Class\,I SEDs, but the dip at $\sim$ 14 $\mu$m will produce a blue 
$\alpha_{\rm IR}^{7-14}$ index. 


\citet{hur96} divided the flux from their HIRES map of IRAS 18272+0114 
between the three sources DEOS, EC53 and S68N, while ISOCAM resolves six 
mid-IR excess sources (cf Fig.~\ref{fig-iram}), of which only EC37 has 
$\alpha_{\rm IR}^{7-14} > 0$. 

It is possible that a broad silicate absorption feature and/or the presence 
of H$_2$O and CO$_2$ ices in the 14.3 $\mu$m band \citep{whi96,deg96} may 
cause absorption effects that correspond in magnitude to the observed dips 
in the SEDs at $\sim$ 14 $\mu$m, noting that the effect is there for the 
most deeply embedded sources. Recently, ISOCAM-CVF spectroscopy of 42 
Class~I and Class~II YSOs in Serpens, $\rho$ Ophiuchi, Chamaeleon and RCrA 
by \citet{ale03} reveals a number of absorption features from 5 to 18 
$\mu$m. They find that the majority of Class~I sources fall in their 
category {\bf a}, having deep absorption features of ices and silicates, 
and conclude that in the cases of large extinction the continuum spectral 
index between 2 and 14 $\mu$m provides a truer value of the shape of the 
underlying continuum than the observed mid-IR spectrum. 
For the sub-sample of 20 sources that were studied by \citet{ale03} in
Serpens we found: four Class~Is, six flat-spectrum sources, four Class~II 
sources, one ``blue'' source, and we did not detect five of their sources. 
All four Class~Is, three flat-spectrum sources, and one Class~II fall in 
their group {\bf a}, and two flat-spectrum sources and three Class~IIs in 
their category {\bf b} (only SVS2 is of type {\bf c}). This comparison shows 
overall good agreement, and indicates that we have not overestimated the 
number of Class~Is in our study.

Also, the presence of shocked molecular H$_2$ line emission in the 6.7 
$\mu$m band \citep[e.g.][]{cab98,lar02} included in the measured flux of some 
of the youngest sources could contribute to give a bluer $\alpha_{\rm IR}^{7-14}$ 
index than expected from a dust continuum. 

Recent 2-D models of Class~I source geometries by \citet{whi03} which include 
flared disk and bipolar cavity produce mid-IR SEDs with very broad dips around 
10 $\mu$m, and overall much bluer mid-IR colours than produced by simple 1D
or simplified 2-D models.


\begin{figure}
\resizebox{\hsize}{!}{\includegraphics{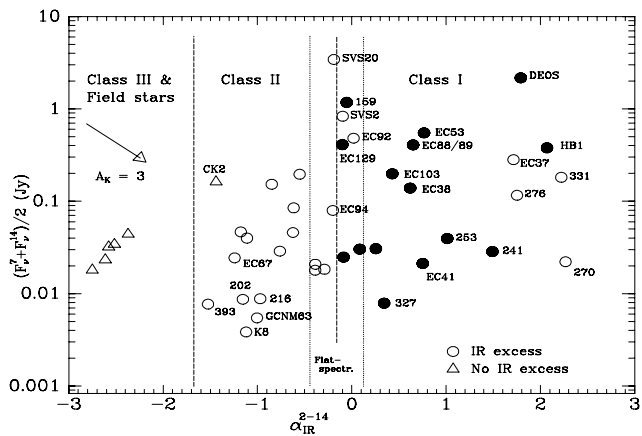}}
\caption{The average flux at 6.7 and 14.3 $\mu$m versus 
         $\alpha_{\rm IR}^{2-14}$ 
         for a subset of the ISOCAM sample. The location of Class I, Class II
         and Class III sources and field stars are indicated. Filled circles
         mark the 15 YSOs with $\alpha_{\rm IR}^{2-14} > 0$ and  
         $\alpha_{\rm IR}^{7-14} < 0$. \label{fig-fav}}
\end{figure}

Based on the previous discussion we trust that the $\alpha_{\rm IR}^{2-14}$ 
index gives a more reliable measure of the SED than the $\alpha_{\rm IR}^{7-14}$ 
index. We have also looked at the mid-IR fluxes of Class~Is versus Class~IIs. 
If all the 15 YSOs with 
$\alpha_{\rm IR}^{2-14} > 0$ and $\alpha_{\rm IR}^{7-14} < 0$ were reddened 
Class\,II sources, one would expect them to be on the average fainter than 
(or at most as bright as) the YSOs with $\alpha_{\rm IR}^{2-14} < 0$. 
We note, however, that there is a slight tendency that sources with large 
$\alpha_{\rm IR}^{2-14}$ indices also are, on the average, the brightest 
ones in the mid-infrared.
Figure~\ref{fig-fav} shows the average flux at 6.7 and 14.3 $\mu$m versus 
$\alpha_{\rm IR}^{2-14}$, with a reddening vector of A$_K = 3$ magnitudes 
inserted. The filled circles in the figure mark the 15 sources that have 
$\alpha_{\rm IR}^{2-14} > 0$ and $\alpha_{\rm IR}^{7-14} < 0$. 
Excluding the flat-spectrum sources, the median fluxes are $0.18$ and $0.029$ 
Jy for {\bf 15} Class\,Is and {\bf 27} Class\,IIs (including all those from 
Table~\ref{tbl-6} that have mid-IR fluxes), respectively. For comparison, 
the median mid-IR fluxes of 16 Class\,Is and 76 Class\,IIs in $\rho$~Ophiuchi 
are $1.17$ and $0.095$ Jy, respectively \citep{bon01}. In Chamaeleon~I for 5 
Class\,Is and 42 Class\,IIs these numbers are $0.28$ and $0.077$ Jy, 
respectively. While the statistics is low for the Chamaeleon~I Class\,I sources,
there is an indication of luminosity evolution from the Class\,I to the 
Class\,II phase, the case being strongest for $\rho$~Ophiuchi. We conclude that
on the basis of this statistical luminosity argument, there seems to be an 
intrinsic difference between the Class\,I and the Class\,II populations.

Noting that Class\,Is are on the average more luminous in the mid-IR than 
Class\,IIs \citep[cf.][]{bon01}, we caution that it is not entirely excluded 
that the large number 
fraction of Class\,I sources in Serpens could be a direct effect of a lower 
sensitivity because of a larger distance. There are about 80 faint YSO 
candidates in the Serpens Cloud Core below the sensitivity limit of ISOCAM 
\citep{kaa99a}.

Although the exact Class\,I/Class\,II number ratio might become subject to 
modification when future observations at high spatial resolution and 
sensitivity towards longer wavelengths are available (SIRTF-Spitzer at 24 
$\mu$m and FIRST-Herschel at 90-250 $\mu$m), we believe that the main 
conclusion of an unusually large fraction of Class\,I sources in the Serpens 
Cloud Core will be maintained.

\subsection{The protostar sample in Serpens}

Based on the discussion in the previous paragraph, we have arrived at a 
sample of 20 Class~I SED sources which are listed in Table~\ref{tbl-4} with 
the two SED indices $\alpha_{\rm IR}^{2-14}$ and $\alpha_{\rm IR}^{2-7}$. 
These are all protostar candidates. Although Class\,0 sources normally
are not expected to be detectable at shorter wavelengths than about 
25 $\mu$m, we detect two Class\,I sources, ISO-241 and ISO-308, within the 
$\pm$ 3$\arcsec$ positional uncertainties of S68N and SMM4, respectively. 
\footnote{According to \citet{and00} the following sources satisfy the Class\,0 
criteria: FIRS1, S68N, SMM3, and SMM4, all of which are bright sources in the 
1.3 mm IRAM map in Fig.~\ref{fig-iram}. In addition, SMM2 is a candidate 
Class\,0. ISO-258b, which may be a knot in the extended emission (cf. 
Fig~\ref{fig-D2}), is 8'' away from FIRS1, but could result from scattered light 
(see Sect.~\ref{ind}). SMM2 and SMM3 have no ISOCAM detections within the 
positional uncertainties, that is to a $3\sigma$ upper limit of 6 and 9 mJy for 
6.7 and 14.3 $\mu$m, respectively (cf. Table~\ref{tbl-1}).} 

\citet{mot01} found that about 60\% of the Class~I sources in Taurus-Auriga 
are ``true protostars'', using the criterion that the envelope mass to 
stellar mass fraction is $\ga 0.1$. A similar method has been applied to the
Serpens data. But first we exclude 3 sources in order to account for the 
possible confusion with Class\,0 sources (in the case of EC41) or contribution 
of Class\,0 sources (in the cases: ISO-241, ISO-308) in the sample of 19 
Class\,Is in the region mapped by IRAM. The fraction of bona-fide protostars 
in the Serpens Core is then estimated to be 9 out of 16 Class\,Is or 56\%, with 
an uncertainty of $\pm$ 10\% because of source confusion problems with the 
strong clustering.


\begin{table}
\caption{The 20 Class~I sources. $\Delta K$ and fwhm refers to
measured variability 1995-1996 \citep{kaa99a} and extendedness in 
the K-band, respectively. K-band magnitudes from 1996 are found in 
Table~\ref{tbl-2}.
         \label{tbl-4}
         }
         
\newcommand\cola {\null}
\newcommand\colb {&}
\newcommand\colc {&}
\newcommand\cold {&}
\newcommand\colf {&}
\newcommand\colg {&}
\newcommand\eol{\\}
\newcommand\extline{&&&&&&\eol}

   \begin{tabular}{rlrrcc}
   \hline
   \noalign{\smallskip}

{ISO} & {Other ID} & $\alpha_{IR}^{2-14}$ & $\alpha_{IR}^{2-7}$ & {fwhm} 
& $\Delta K$ \\
\#    &                  &                 &                & $\arcsec$  
&  mag  \\
\hline
\cola 270\colb \colc  2.56\cold  3.54\colf   2.31\colg  \eol
\cola 331\colb \colc  2.51\cold  4.12\colf   4.57\colg  \eol
\cola 330\colb HB1\colc  2.35\cold  4.26\colf   1.47\colg 0.32\eol
\cola 250\colb DEOS\colc  2.06\cold  3.52\colf   7.52\colg 0.92\eol
\cola 276\colb GCNM53\colc  2.02\cold  3.05\colf   1.89\colg \eol
\cola 249\colb EC37\colc  1.98\cold  2.66\colf   1.25\colg 0.26\eol
\cola 241\colb \colc  1.74\cold  3.26\colf   1.49\colg \eol
\cola 308\colb HCE170/171$^1$\colc -\cold  2.84\colf   4.48\colg 0.64\eol
\cola 253\colb EC40\colc  1.23\cold  2.76\colf   1.21\colg \eol
\cola 265\colb EC53\colc  0.98\cold  2.27\colf   2.27\colg 0.59\eol
\cola 259\colb \colc -\cold 1.94$^2$\colf -\colg \eol
\cola 260\colb \colc -\cold  1.88\colf   1.83\colg \eol
\cola 258a\colb EC41\colc  0.96\cold  1.81\colf   1.33\colg \eol
\cola 312\colb EC88/89$^3$\colc  0.85\cold  1.73\colf   9.53\colg 0.11\eol
\cola 254\colb EC38\colc  0.82\cold  1.78\colf   1.00$^4$\colg -\eol
\cola 326\colb EC103\colc  0.62\cold  1.27\colf   1.55\colg 0.22\eol
\cola 327\colb HCE175\colc  0.53\cold  1.73\colf   1.42\colg \eol
\cola 306\colb EC80\colc  0.44\cold  1.70\colf   1.32\colg \eol
\cola 277\colb EC63\colc -\cold  1.21\colf   1.26\colg \eol
\cola 313\colb GCNM94 \colc -\cold  1.19\colf   1.32\colg 0.43\eol

   \noalign{\smallskip}
   \hline
   \end{tabular}

$^1$ both are extended, HCE171=K32 is variable  \\
$^2$ Ks = 14.48 from 2MASS PSC, 16'' south-west of extended IRAS 18273+0059,
possible K extendedness not investigated \\
$^3$ both are extended, EC88 is variable \\
$^4$ from image serp-nw which has better seeing \\

Identifier acronyms as in Table~\ref{tbl-2}.

\end{table}

Extended, elongated and polarized near-IR emission is predicted by models of 
infalling envelopes which have developed cavities owing to bipolar outflows, 
where near-IR radiation from the central object may escape and scatter off 
dust in the cavity and the outer envelope, see e.g. \citet{whi97}. \citet{par02} 
find that 70\% of the bona-fide protostars in Taurus are extended in the 
near-IR, and that less than 10\% of the sources that are 
extended in the near-IR show no extension in mm continuum. Our near-IR data
has a spatial resolution about 10 times better than the IRAM map, and we have 
included in Table~\ref{tbl-4} the full width at half maximum (fwhm) of a 
gaussian fit to the source profiles in the $K$ band images. The median fwhm 
of 160 isolated and relatively bright sources over the field is $1.\arcsec27 
\pm 0.\arcsec10$. Sources with a fwhm greater than $1.\arcsec6$, i.e. above 
3$\sigma$, are here defined as extended in the K-band. For the Class~I sources 
this concerns 8 of 19 sources or 42\%, highly coinciding with continuum 
sources except for ISO-331 and ISO-260. There are no extended sources in the 
flat-spectrum and Class\,II samples. 

Substantial K-band variability was found in 8 of the above Class~I sources, 
5 of the flat-spectrum sources and 5 of the Class~II sources \citep{kaa99a}. 
The brightness variations of the Class~I sources are given in Table~\ref{tbl-4} 
as $\Delta K$. The median value of the amplitude is $0.38 \pm 0.27$ mag, 
larger than the variations found for Class~Is in Taurus \citep{par02}.


\begin{table}
\caption{The 13 flat-spectrum sources. $\Delta K$ and fwhm refers to
measured variability 1995-1996 \citep{kaa99a} and extendedness in 
the K-band, respectively. K-band magnitudes from 1996 are found in 
Table~\ref{tbl-1}.
         \label{tbl-5}
         }
         
\newcommand\cola {\null}
\newcommand\colb {&}
\newcommand\colc {&}
\newcommand\cold {&}
\newcommand\cole {&}
\newcommand\colf {&}
\newcommand\eol{\\}
\newcommand\extline{&&&&&&\eol}

   \begin{tabular}{rlrrcr}
   \hline
   \noalign{\smallskip}

{ISO} & {Other ID} & $\alpha_{IR}^{2-14}$ & $\alpha_{IR}^{2-7}$ & {fwhm} 
& $\Delta K$ \\
\#    &                  &                      &                     
& $\arcsec$ & mag \\
\hline
\cola 234\colb EC26\colc -\cold  0.63\cole   1.31\colf \eol
\cola 322\colb EC98\colc -\cold  0.17\cole   1.31\colf \eol
\cola 345\colb EC125\colc  0.26\cold  0.44\cole   1.30\colf \eol
\cola 317\colb EC92/95\colc  0.19\cold -0.03\cole   1.35\colf \eol
\cola 159\colb IRAS 18269+0116\colc  0.11\cold  0.23\cole   1.00$^1$\colf -\eol
\cola 237\colb EC28\colc  0.07\cold  0.99\cole   1.23\colf \eol
\cola 307\colb SVS2\colc  0.06\cold -0.06\cole   1.38\colf  0.19\eol
\cola 347\colb EC129\colc  0.06\cold  0.53\cole   1.34\colf \eol
\cola 314\colb SVS20 (double)\colc -0.03\cold  0.30\cole   2.74$^2$\colf  0.14\eol
\cola 318\colb EC94\colc -0.05\cold  0.56\cole   1.35\colf \eol
\cola 341\colb EC121\colc -0.14\cold  0.30\cole   1.27\colf  0.21\eol
\cola 294\colb EC73\colc -0.24\cold -0.18\cole   1.31\colf  0.66\eol
\cola 320\colb EC91\colc -0.24\cold  0.45\cole   1.33\colf  0.32\eol
    \noalign{\smallskip}
    \hline
    \end{tabular}

$^1$ From image serp-nw which has better seeing. \\
$^2$ Double source, not extended. \\

Identifier acronyms as in Table~\ref{tbl-2}.

\end{table}

The flat-spectrum sources are believed to be in a transition phase between 
the Class~I and the Class~II stage, but a few of these might be found to be 
true protostars. We have listed them in Table~\ref{tbl-5} with SED indices, 
$K$-band fwhm and variability amplitude. Both SVS2 and SVS20 have 
centrosymmetric polarization patterns, indicating that evacuated bipolar 
cavities surround them, and both are double sources \citep{hua97}. 
Among the flat-spectrum sources ISO-159, EC129, SVS20, and ISO-237 (i.e. 
30\%) are detected as mm continuum sources in our IRAM map.


\begin{table*}
\caption{The 43 ISOCAM YSOs found to be Class~II sources or candidates. 
A hyphen for the SED indices means no detection at 14 $\mu$m. $A_K$ is 
the derived extinction in the K-band, and $L_{\star}$ is the stellar 
luminosity estimated as explained in Sect.~\ref{lf}. $\Delta K$ is the 
amplitude of the K-band variability 1995-1996 \citep{kaa99a}. None of 
the sources for which Arnica $K$ images were available was found to be 
extended in $K$. 
         \label{tbl-6}
         }

\newcommand\cola {\null}
\newcommand\colb {&}
\newcommand\colc {&}
\newcommand\cold {&}
\newcommand\cole {&}
\newcommand\colf {&}
\newcommand\colg {&}
\newcommand\eol{\\}
\newcommand\extline{&&&&&&\eol}

\begin{tabular}{rlccrrl}
\hline
\noalign{\smallskip}

{ISO} & {Other ID} & $\alpha_{IR}^{2-14}$ & $\alpha_{IR}^{2-7}$ & $A_K$ &
$L_{\star}$ & {Additional information} \\ 
\#    &                  &                      &                &
 (mag)  &   ($L_{\odot}$)                &          \\
\hline
\cola  29\colb \colc -2.13\cold -3.96\cole 0.32 \colf 0.009
\colf Mid-IR excess, $Ks = 11.17$ from 2MASS \eol
\cola 150\colb \colc -1.30\cold -1.59\cole 0.30 \colf 0.19
\colg 15'' north of radio source S68 3, $Ks = 10.69$ from 2MASS \eol
\cola 158\colb BD+01 3687\colc -1.88\cold -2.92 \cole 0.0 \colf 0.086
\colg $K = 9.8$ from \citet{cha88}. Transition object II-III?\eol
\cola 160\colb \colc -\cold -0.69 \cole 2.28 \colf 2.4
\colg See Appendix \ref{ind} \eol
\cola 173\colb [CDF88] 6\colc -1.58 \cold -2.46 \cole 0.0 \colf 0.57 
\colg $Ks = 9.21$ from 2MASS \eol
\cola 202\colb \colc -1.05\cold -1.42 \cole 0.77 \colf 0.096
\colg Located 30'' south-east of HH460\eol
\cola 207\colb STGM 3\colc -1.18 \cold -1.48 \cole 0.35 \colf 0.57
\colg $Ks = 9.64$ from 2MASS, NIR excess \citep{sog97} \eol
\cola 216\colb \colc -0.86\cold -1.47 \cole 0.0 \colf 0.057
\colg Located 1' east of HH477\eol
\cola 219\colb STGM 2\colc -\cold -1.70 \cole 0.58 \colf 0.077
\colg $Ks = 11.60$ from 2MASS, NIR excess \citep{sog97} \eol
\cola 221\colb IRAS 18271+0102\colc -1.14\cold -1.53\cole 1.09 \colf 36
\colg $Ks = 5.38$ from 2MASS, see also \citet{cla91} \eol
\cola 224\colb EC11\colc -\cold -1.58 \cole 0.11 \colf 0.062
\colg Clear IR excess in the $H-K/K-m_7$ diagram \eol
\cola 226\colb EC13\colc -\cold -0.84 \cole $\sim$ 2.0 \colf 0.061
\colg Clear IR excess in the $H-K/K-m_7$ diagram\eol
\cola 231\colb EC21\colc -\cold -1.18 \cole 3.30 \colf 0.33
\colg Uncertain, see Appendix \ref{ind}\eol
\cola 232\colb EC23\colc -\cold -2.33 \cole 0.69 \colf 0.072
\colg Uncertain, See Appendix \ref{ind} \eol
\cola 242\colb K8,EC33\colc -1.02\cold -0.80 \cole 2.28 \colf 0.099
\colg Variable, $\Delta K=0.21$ \eol
\cola 252\colb \colc -0.87 \cold -1.15\cole 0.70 \colf 0.24
\colg $Ks = 11.12$ from 2MASS \eol
\cola 266\colb GCNM35,EC51\colc -\cold -1.77 \cole 0.76 \colf 0.021
\colg NIR excess\eol
\cola 269\colb K16,EC56\colc -\cold -1.58 \cole $\sim$ 1.7 \colf 0.029
\colg Variable, $\Delta K=0.21$ \eol
\cola 272\colb EC59\colc -\cold -1.82 \cole 3.42 \colf 0.086
\colg Uncertain, see App.\ref{ind} \eol
\cola 279\colb STGM14,EC66\colc -\cold -1.56 \cole $\sim$ 1.4 \colf 0.020 
\colg Variable, $\Delta K=0.21$ + NIR excess\eol
\cola 283\colb EC67\colc -1.14\cold -1.22 \cole $\sim$ 0.51 \colf 0.32
\colg \eol
\cola 285\colb GCNM63,EC68\colc -0.89\cold -0.77 \cole 1.47 \colf 0.097
\colg \eol
\cola 287\colb \colc -0.80\cold -1.04\cole 0.75 \colf 0.030
\colg Mid-IR excess + optically visible, likely Class\,II \eol
\cola 289\colb EC69,CK10\colc -\cold -1.42 \cole 2.56 \colf 0.45
\colg See Appendix \ref{ind} \eol
\cola 291\colb GCNM70,EC70\colc -\cold -2.05 \cole 0.09 \colf 0.016
\colg NIR excess\eol
\cola 298\colb EC74,CK9,GEL4\colc -0.48\cold -0.29 \cole 1.67 \colf 1.7
\colg Variable, $\Delta K=0.15$ \eol
\cola 304\colb EC79,GEL5\colc -0.64\cold -0.52 \cole 1.42 \colf 0.50
\colg \eol
\cola 309\colb EC84,GEL7\colc -0.49\cold -0.55 \cole 1.80 \colf 0.79
\colg \eol
\cola 319\colb EC93,CK13\colc -\cold -0.89 \cole 2.35 \colf 0.97
\colg Clear IR excess in the $H-K/K-m_7$ diagram \eol
\cola 321\colb CK4,GEL12,EC97\colc -0.42\cold -0.44 \cole 1.16 \colf 2.8
\colg \eol
\cola 328\colb EC105,CK8,GEL13\colc -0.73\cold -0.38 \cole 0.55 \colf 2.4
\colg Variable, $\Delta K=0.63$\eol
\cola 338\colb EC117,CK6,GEL15\colc -\cold -2.21 \cole 1.36 \colf 0.25
\colg No IR excess, but ass. with nebulosity \eol
\cola 348\colb EC135,GGD29\colc -1.01\cold -1.34 \cole 0.39 \colf 0.42
\colg \eol
\cola 351\colb EC141\colc -\cold -1.12 \cole 2.05 \colf 0.16
\colg Clear IR excess in the $H-K/K-m_7$ diagram\eol
\cola 356\colb K40,EC152\colc -\cold -1.03 \cole 0.70 \colf 0.030
\colg Clear IR excess in the $H-K/K-m_7$ diagram \eol
\cola 357\colb \colc -1.11\cold -1.73\cole 0.25 \colf 0.096
\colg $Ks = 11.23$ from 2MASS \eol
\cola 359\colb \colc -0.92\cold -1.20\cole 0.38 \colf 0.33
\colg $Ks = 10.56$ from 2MASS, strong H$\alpha$ emission (to be publ.)\eol
\cola 366\colb STGM8\colc -1.08\cold -1.29 \cole 0.59 \colf 0.59
\colg \eol
\cola 367\colb \colc -0.87\cold -1.16\cole 0.34 \colf 0.12
\colg $Ks = 11.67$ from 2MASS, strong H$\alpha$ em.\eol
\cola 370\colb \colc -0.83\cold -1.00\cole 0.02 \colf 0.54
\colg $Ks = 10.13$ from 2MASS, dbl in the optical, both H$\alpha$ em. 
(to be publ.)  \eol
\cola 379\colb \colc -1.95\cold -2.24\cole 0.81 \colf 1.2
\colg $Ks = 8.09$ from 2MASS \eol
\cola 393\colb \colc -1.44\cold -2.27 \cole 0.44 \colf 0.066
\colg double in $K$ \eol
\cola 407\colb \colc -1.24\cold -2.13\cole 0.19 \colf 0.021
\colg $Ks = 12.36$ from 2MASS \eol

\noalign{\smallskip}
\hline
\end{tabular}

         Identifier acronyms as in Tab~\ref{tbl-2}.
\end{table*}

\subsection{The pre-main sequence sample in Serpens}

Based on the ISOCAM YSO sample found in Table~\ref{tbl-1} we have used a 
combination of criteria to arrive at a tentative sample of Class\,II sources 
in Serpens. These are listed in Table~\ref{tbl-6} with SED indices, whenever 
available, and additional criteria used to argue for a Class\,II status. Since 
our $K$ imaging from 1996 only covers about 10\% of the ISOCAM survey, we 
have also used published $K$ band photometry whenever available to calculate 
the SED indices. 
Also, we have assumed that if the source shows strong mid-IR excess and is 
optically visible, it is more likely to be a Class\,II than a flat-spectrum
or Class\,I type of YSO. As a supporting argument for Class\,II designation 
we have found for a few mid-IR excess sources without JHK data, strong 
H$\alpha$ in emission (to be published in a future paper). When the 2MASS
All-Sky Data Release became available while we were finalizing our 
investigation,
we used the near-IR photometry from their point source catalogue for the 
Class\,II source candidates where such data was lacking, and found spectral 
indices in agreement with our above reasoning.

The pre-main sequence population also includes the Class\,III type of YSOs, 
objects which show no IR signatures of an optically thick disk. Since the
basic criterion for selecting YSOs in our study is IR excess, we do not
sample these sources. With the existing photometric data alone we cannot 
distinguish Class\,IIIs from field stars, except for a few objects 
in the transition zone between the Class\,II and Class\,III phases. The 
clear gap between the two groups formed in the ISOCAM colour-magnitude 
diagram in Fig.~\ref{fig-323} demonstrates that there are few objects 
undergoing this transition, i.e. that the transition must be rapid. This
is evident in all the star formation regions surveyed by ISOCAM. In
Serpens at least five sources from Table~\ref{tbl-3} belong to this 
transition group, cf. Fig.~\ref{fig-323}, and we list them as candidate 
Class\,IIIs in Table~\ref{tbl-7}. None of these are located in the central 
Cloud Core region covered by Arnica deep near-IR imaging.


\begin{table}
\caption{Five candidate Class\,III sources from Table~\ref{tbl-3}. These are
         probably in the transition phase between Class\,II and III having 
         some remaining IR excess.
         K-band magnitudes for two sources are obtained from \citet{kaa95}. 
         \label{tbl-7}
         }

\newcommand\cola {\null}
\newcommand\colb {&}
\newcommand\colc {&}
\newcommand\cold {&}
\newcommand\cole {&}
\newcommand\eol{\\}
\newcommand\extline{&&&&\eol}

\begin{tabular}{rlrrr}
\hline
\noalign{\smallskip}

{ISO} & {Other ID} & $\alpha_{IR}^{7-14}$ & $\alpha_{IR}^{2-14}$ &
$\alpha_{IR}^{2-7}$ \\
\#    &            &                      &                      &
                     \\     
\hline
\cola 112\colb \colc -1.67\cold  \cole  \eol
\cola 190\colb \colc -1.98\cold -1.83\cole -1.73\eol
\cola 210\colb \colc -2.00\cold   \cole \eol
\cola 311\colb BD+01 3689B\colc -1.73\cold -2.21\cole -2.54\eol
\cola 332\colb \colc -1.72\cold  \cole \eol
\noalign{\smallskip}
\hline
\end{tabular}

Identifier acronyms as in Table~\ref{tbl-3}.

\end{table}


\section{Luminosity distribution}
\label{lf}

\begin{figure}
\resizebox{\hsize}{!}{\includegraphics{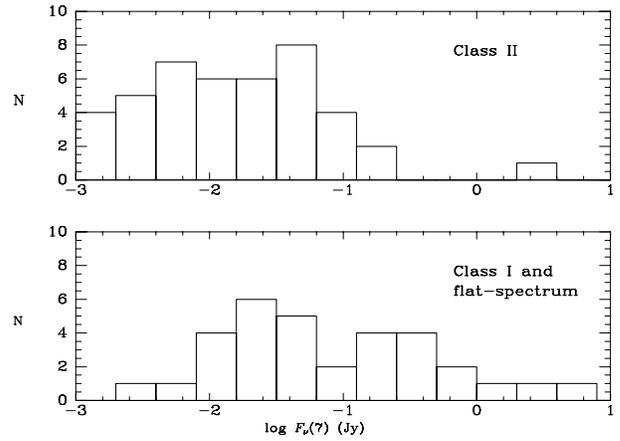}}
\caption{The 6.7 $\mu$m luminosity function for the sample of Class\,Is and 
flat-spectrum sources (lower) and Class\,II sources (upper). 
          \label{fig-lf}}
\end{figure}

Only about 10\% of the Serpens area surveyed by ISOCAM was covered by deep
JHK photometry. The total sample is therefore not as homogeneous as 
our surveys in $\rho$ Ophiuchi \citep{bon01}, Chamaeleon \citep{per00} and 
RCrA \citep{olo99}. Also, since Serpens is at a larger distance the overall
sensitivity is lower, and source confusion is worse, especially in the very
clustered active centre. With our deep RIJHK photometric
coverage over the whole ISOCAM survey, which will be published in a future 
paper, we expect to extend the population of IR excess 
sources, and reveal YSOs which are below the ISO sensitivity. With near-IR 
photometry for the whole sample we can also improve the luminosity
estimate of the Class\,II sources, by dereddening the J-band fluxes, 
analogously to what was done for $\rho$ Ophiuchi \citep{bon01}. 
At the moment we estimate the luminosities from the mid-IR 6.7 $\mu$m band
only, because this is the most homogeneous measurement we have. The 6.7 
$\mu$m luminosity function is shown in Fig.~\ref{fig-lf} for the Class\,I 
and flat-spectrum sources (lower) and Class\,II sources (upper). 

Protostars are expected to radiate most of their luminosity at longer 
wavelengths. To derive their bolometric luminosities we need to integrate 
the observed SEDs. As shown in Sect.~\ref{clustering}, however, the four 
IRAS sources found in the CE region correspond each to a number of protostars. 
For reasons of source confusion we have not attempted to make calorimetric 
luminosity estimates of the Class\,Is in Serpens, as was done for the 
Class\,I sample in $\rho$ Ophiuchi, see \citet{bon01} who 
found that the typical fraction of the luminosity radiated between 6.7 
and 14.3 $\mu$m for a Class\,I source is $\sim 10 \%$. 

As shown in Fig.~\ref{fig-lf} the protostars span a range in mid-IR 
luminosities about equal to that of the Class\,IIs. It is clear that the 
protostars are on the average more luminous at 6.7 $\mu$m than the 
Class\,IIs.  This is expected as a large fraction of the luminosity of a 
protostar is the accretion luminosity. The accretion can be partly 
continuous and partly happen in bursts, so that the luminosity of a 
protostar is far from being a simple function of its age and mass. 

Class\,II sources are found to radiate the bulk of their flux in the 
near-IR. The 6.7 $\mu$m band is expected to be dominated by disk emission, 
e.g. for a simple black-body at T = 3700 K only about 8\% of the luminosity 
is radiated in this band \citep{bon01}. Thus we can relate the mid-IR flux 
to the total stellar 
luminosity by assuming that what we see is dust emission from a passive 
reprocessing circumstellar disk. For most Class\,II sources any contribution 
to the luminosity from an active accretion disk is assumed to be negligible. 
This is supported by measurements of the disk accretion rates of CTTS in 
Taurus, which are found to have a median value of only $10^{-8}$ M$_{\odot}$ 
yr$^{-1}$ \citep[e.g.][]{gul98}. If the dust in the disk is distributed 
roughly as in the ideal case of an infinite, spatially flat disk, 25 \% of 
the central source luminosity is absorbed and reradiated by the dust 
\citep{ada86}. If the disk is flared, the percentage increases, and if it has 
an inner hole, it decreases. Furthermore, the disk inclination angle $i$ (to 
the line of sight) determines what fraction of the disk luminosity we observe.
In addition, we have to correct for cloud extinction. Since many of our 
Class\,II sources are not detected at 14.3$\mu$m we select to use the 
6.7$\mu$m flux only. 


\citet{olo99} calibrated an empirical relation between the observed
6.7$\mu$m flux and the stellar luminosity by selecting sources with known
spectral class and very small extinction in RCrA and Chamaeleon I. 
Correcting this relation to the larger distance of Serpens we get here
that $\log (L_{\star}/L_{\odot}) = \log F_{6.7} (Jy) + 0.95$. Before we 
can apply this relation, however, we have to correct $F_{6.7}$ for cloud 
extinction. 
For the 21 Class\,IIs which have Arnica JHK photometry, we assume an intrinsic 
colour of $(J-H)_0 = 0.85$, which is a median value with small dispersion 
found for CTTS \citep{str93,mey97}, and estimate the extinction in the 
K-band, $A_K = 0.97 \times [(J-H) - (J-H)_0]$, applying the $\lambda^{-1.9}$ 
extinction law for the near-IR found for Serpens \citep{kaa99a}. Sources 
without J-band photometry are either not detected in $J$ - probably owing 
to extinction - or they are located outside the area observed in the near-IR. 
In the first case (ISO-Ser-226, 269, 279) we interpolate extinction values 
of their neighbouring Class\,II sources, and in the second case we use the
recently available 2MASS PSC and a dereddening as above. We do not transform
from 2MASS $Ks$ to Arnica $K$, since the difference is less than or about 
equal to the estimated errors in the photometry (0.01-0.02 mag).

All extinction values are listed in Table~\ref{tbl-6} expressed as $A_K$, 
but only approximate values are given for ISO-Ser-226, 269 and 279, since 
it cannot be known at which depth in the cloud these objects reside. In 
Sect.~\ref{nir} we found the relation $A_{6.7} = 0.41 A_K$, which is used
to correct the 6.7 $\mu$m flux. The derived stellar luminosity $L_{\star}$
for each Class\,II source is also listed in Table~\ref{tbl-6}. The uncertainty 
in the luminosity estimate is a function of the uncertainties in: 1) the 
extinction estimate, 2) the distance, 3) the $L_{\star}$ vs. $F_{6.7}$ relation 
itself, of which the last two
will be systematic errors for the whole sample. Comparison with previous
luminosity estimates by \citet{eir92}, but scaled to the distance we use, 
shows that for the 13 objects that overlap in these two samples, there is
a large scatter; the fraction $L/L_{EC92}$ varies from 0.5 to 3.7 but the 
median value is 1.1.

The completeness estimate we found in Sect.~\ref{stat} for detections in
the 6.7 $\mu$m band is 6 mJy. Because of the variable extinction this cannot 
be directly translated to a completeness for Class\,II sources in terms 
of stellar luminosities. We have used the average measured extinction of 
$A_K = 1.0$ for our Class\,II sample and calculated that the completeness 
limit is at $L = 0.08 L_{\odot}$. 

\section{Implications for the IMF in Serpens}
\label{imf}

\citet{eir92} first estimated a luminosity function (LF) for the young cluster
in the Serpens Cloud Core, including their 51 identified cluster members. The 
stellar luminosities were obtained from a trapezoidal integration over the
detected wavelengths, extrapolated to longer wavelengths and 
extinction-corrected. This LF showed a pronounced peak around 1 $L_{\odot}$ and a
turnover below 0.2 $L_{\odot}$. \citet{gio98} evaluated synthetic K-band
luminosity functions (KLFs) and found a best fit to their observed KLF with 
two bursts of star formation, one 0.1 Myrs ago and the other around 3 Myrs ago, 
and an underlying Miller \& Scalo IMF. This was partly based on their finding 
of a turnover in the KLF above $K = 14$ mag. A later study expanded the number 
of Serpens members and found no evidence for a turnover of the KLF down to a 
limit of $K = 16$ mag \citep{kaa99a}. The weakness of KLFs is that differential 
extinction is not taken into account, and for Serpens this is especially 
important as values as high as $A_V \sim 50$ mag have been observed for background 
stars, and $A_V \sim 100$ mag is expected for the densest regions of the NW-SE 
ridge \citep{whi95}.

\begin{figure}
\resizebox{\hsize}{!}{\includegraphics{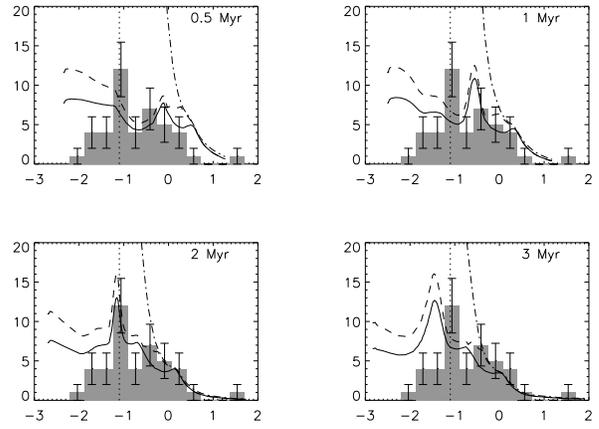}}
\caption{Synthetic LFs were calculated for 4 different ages of a coeval 
population on the basis of pre-main sequence tracks of \citet{dan98} and 3 
different underlying IMFs. The number of sources per bin is plotted vs. 
$\log L$, and the synthetic LFs are plotted over the observed LF for the 
Class\,II sources in Serpens (histogram). Both the \citet{sca98} 
three-segment power-law IMF (solid) and the \citet{kro93} three-segment 
power-law IMF (dashed) seem plausible for a coeval population of $\sim$ 
2 Myr all the way down to the estimated completeness limit at $L \sim 
0.08 L_{\odot}$ (dotted vertical line). The observed LF is also consistent 
with the Salpeter IMF (dashed-dotted) for this age at the high luminosity 
end (it has been extrapolated to low masses for reference only).
\label{fig-lfimf}}
\end{figure}

By selecting sources in the Class\,II phase only, we here constrain the age
spread somewhat \citep{bon01}. This phase is thought to last of the order 
of a few Myr. This means that Class\,IIs can have a similar age spread, but we
cannot rule out that the Class\,IIs once formed in a burst like we see now
for the protostars (cf. Sect.~\ref{clustering}). We have here made the simple 
assumption of coeval star formation. In Fig.~\ref{fig-lfimf} we show our 
observed LF for the sample of 43 Class\,II sources (shaded histograms). 
The bin width of the histogram is $d \log L = 0.325$, based on a factor of 
two uncertainty in the luminosity estimate, and the histogram has been shifted 
and the bin size slightly varied to check that the presented distribution 
remains stable. The observed LF shows a pronounced peak at $L \sim 0.09 
L_{\odot}$, but this corresponds roughly to the completeness limit we have 
estimated for this sample (dotted vertical line). The error bars are the 
statistical counting errors ($\sqrt N$). The many luminous objects of around 
$ 1 L_{\odot}$ found by \citet{eir92} have disappeared as the sample has been
restricted to Class\,II sources. 


We have assumed coeval populations and three different underlying IMFs: 
the Salpeter IMF \citep{sal55}, the Kroupa, Tout and Gilmore three-segment 
power-law IMF \citep{kro93}, and the Scalo three-segment power-law IMF 
\citep{sca98}, hereafter S55, KTG93 and S98, respectively. With the IMF in 
the form: $d N/d \log M_{\star} \propto M_{\star}^{\alpha}$, the 
Salpeter IMF has one single index: $\alpha = -1.35$, originally determined
for the mass interval 0.4 to 10 $M_{\odot}$ but in Fig.~\ref{fig-lfimf} 
extrapolated to lower masses for reference. Both the KTG93 and S98 IMFs 
have three different values of $\alpha$ for three (differently divided) 
segments. \citet{bon01} found for a large sample of Class\,II objects in 
$\rho$ Ophiuchi that the high mass end of the mass function was well fitted 
with the index $\alpha = -1.7$. 
Also, they found that the IMF starts to flatten at $M_{flat} = 0.55 M_{\odot}$ 
and stays ``flat'' down to $0.055 M_{\odot}$ with a power-law index 
$\alpha_{flat} = -0.35$. These best fits of the two free parameters $M_{flat}$ 
and $\alpha_{flat}$ result in a mass function close to both the KTG93 and S98 
IMFs for $\rho$ Ophiuchi. 


\begin{figure*}[htbp]
\resizebox{\hsize}{!}{\includegraphics{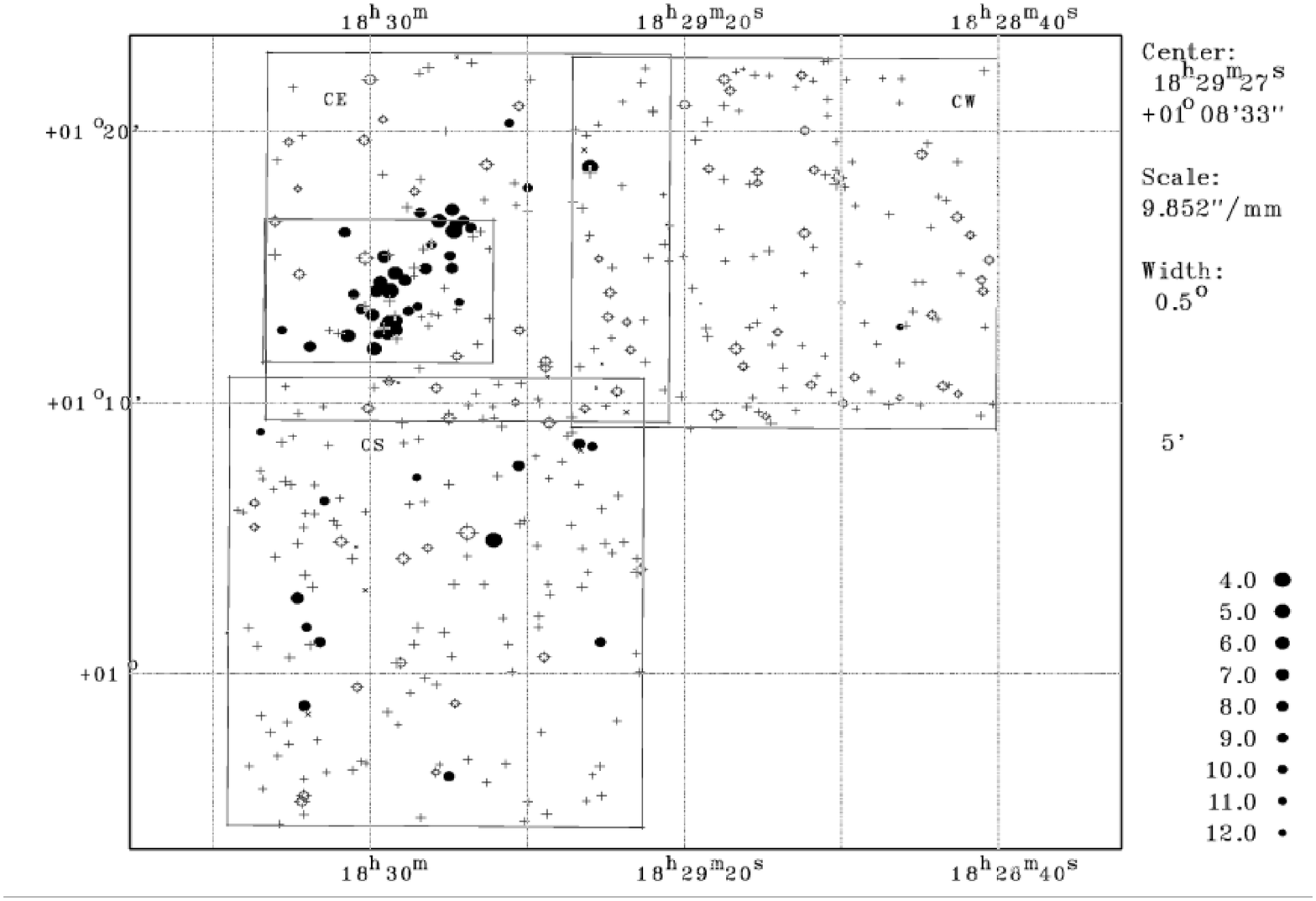}}
\caption{Spatial distribution of the ISOCAM sources in the Serpens Cloud.
         Sources with mid-IR excesses (filled circles) are seen to cluster 
         strongly in the Cloud Core (field CE), while the CW field contains 
         almost exclusively sources without mid-IR excesses (open circles). 
         Indicated are also sources with 6.7 $\mu$m detections only (plus 
         signs) and 14.3 $\mu$m detections only (crosses). The coordinates 
         have epoch J2000. Each of the 3 separate ISOCAM rasters, CE, CW, 
         and CS are indicated, and the small box outlines the region where 
         deep JHK imaging is available \citep{kaa99a}.
\label{figmap}}
\end{figure*}

On the basis of the pre-main sequence evolutionary models of \citet{dan94,dan98}, 
with the 1998 upgrade for low mass stars, we have computed synthetic 
LFs for each IMF and for the four different ages: 0.5, 1, 2 and 3 Myr of a 
coeval cluster. Each window in Fig.~\ref{fig-lfimf} shows the result for 
one age, and the computed LFs are overplotted on the observed LF. 
The peak in the LF which wanders towards lower luminosities with increasing 
age arises because of the deuterium burning phase \citep{zin93}. Deuterium 
burning acts like a thermostat and hampers somewhat the contraction down the 
Hayashi track, causing a build-up of sources in a given luminosity bin for 
the case of coeval star formation. The peak in the observed LF, however, is 
approximately at the completeness limit of the sample, and we cannot put too 
much confidence in it. Nevertheless, it is obvious from the figure that the 
observed LF excludes ages of less than about 1 Myr for this Class\,II 
population for any of three underlying IMFs, which is in agreement with current 
understanding of YSO evolution. It is not possible to distinguish between the 
S98 IMF (solid line) and the KTG93 IMF (dashed), but both of them are plausible 
for a coeval population with an age of about 2 Myr, all the way down to the
estimated completeness limit. Also the Salpeter IMF (dashed-dotted line) is in 
agreement with the observed LF for ages of 2-3 Myr at the high luminosity end. 
Its extrapolation to lower masses is given in the plots just for reference.
 
Note, however, that if star formation proceeds in a more continuous fashion, the 
structures in the model LFs will smooth out. We have checked the model LFs 
for the case of continuous star formation over several intervals from 
0.5 to 3 Myrs. Although the statistics in the observed LF is poor, it
seems we can exclude the combination of the above tested IMFs and 
continuous SF over periods of less than 2 Myrs. We also tested the case 
of a single star formation burst taking place 2 Myr ago but having various 
durations from 0.2 to 1.0 Myr. The peak in the observed LF is not secure
enough to discriminate between the various burst durations. For the tested 
IMFs and SF histories, all we can conclude is that an age of less than 2 Myr 
for the Class II population seems implausible.

Corrections for binarity have not been applied, but such a correction 
would increase the populations in the lowest luminosity bins. We conclude 
that with the currently obtained LF the Class\,II mass function in Serpens is 
similar to the one in $\rho$ Ophiuchi \citep{bon01}, even though flat-spectrum 
sources were included in the $\rho$ Ophiuchi sample. We note that the typical 
stellar mass found for the Serpens Class\,II sample (i.e. the median) is 0.17 
$M_{\odot}$, the same as in $\rho$ Ophiuchi \citep{bon01}. The total mass of 
the Class\,II sample in Serpens is $16.3 M_{\odot}$.

We have also used the \citet{bar98} evolutionary tracks and calculated
synthetic LFs with the same underlying IMFs as above and for the ages:
1, 2, 3, and 4 Myr. This model gives basically the same result as that of
\citet{dan98} for the degree of confidence we can put in our observed LF, 
taking the large error bars in the histogram into account.

Adopting the above deduced age of 2 Myr for the Serpens Class\,II 
population, our study is estimated to be complete to about $0.17 M_{\odot}$, 
but to reach down to $0.03-0.04 M_{\odot}$. Taking $0.08 M_{\odot}$ as the 
border between low mass stars and brown dwarfs, we find 9 sub-stellar size 
objects in the Serpens Class\,II sample (those with $L_{\star} \le 0.04 
L_{\odot}$ in Tab.~\ref{tbl-6}). These are young {\em brown dwarfs} 
(BDs) apparently going through a Class\,II phase just like low mass 
stars. They all have IR excesses with no apparent deviations from other 
Class\,IIs in any aspect, which is consistent with the idea that they formed
in the same way as low-mass stars do. Our sample of BDs is not complete, 
however, and we cannot discuss the frequency of occurrence of disks among
BDs or set strong constraints on BD formation models.

Based on the age estimate above, we find that 
$\sim$ 21\% of the Class\,II sample studied is made up of free-floating 
brown dwarfs. This is similar to the percentage found in $\rho$ Ophiuchi 
\citep{bon01}. The percentage in terms of mass is, however, only about 
3\% of the total mass for the whole Class\,II sample.

\section{Spatial distributions}
\label{clustering}

The spatial distribution of the ISOCAM sources is shown in Fig.~\ref{figmap} 
together with the location of the three main rasters CE, CW and CS. The filled 
circles indicate sources with mid-IR excesses (cf. Table~\ref{tbl-2}), the 
open circles are the sources without mid-IR excesses (cf. Table~\ref{tbl-3}), 
the plus signs are sources with 6.7 $\mu$m fluxes only, and crosses sources 
with 14.3 $\mu$m fluxes only. The small box outlines the $8\arcmin \times 
6\arcmin$ field for which deep JHK imaging is obtained. The overall spatial 
distribution of the mid-IR excess YSOs shows a very clear concentration along 
a SE-NW oriented ridge within the CE field, mainly coinciding with what is 
known as the Serpens Cloud Core. In the CW field there is practically no sign 
of activity, while the distribution of IR-excess YSOs in the CS field is quite 
scattered. The majority of the mid-IR excess sources seem to be distributed 
along a curved filament pointing towards a centre of curvature around 18:29:10, 
+01:05:00 (J2000). Nothing particular was found at this location searching
SIMBAD, however, and whether star formation along this arc may be externally 
triggered remains speculative. The IRAS Point Source Catalogue contains six 
sources in the area surveyed by ISOCAM (cf. Appendix~\ref{ind} for details). 


\begin{figure*}[htbp]
\resizebox{\hsize}{!}{\rotatebox{0}{\includegraphics{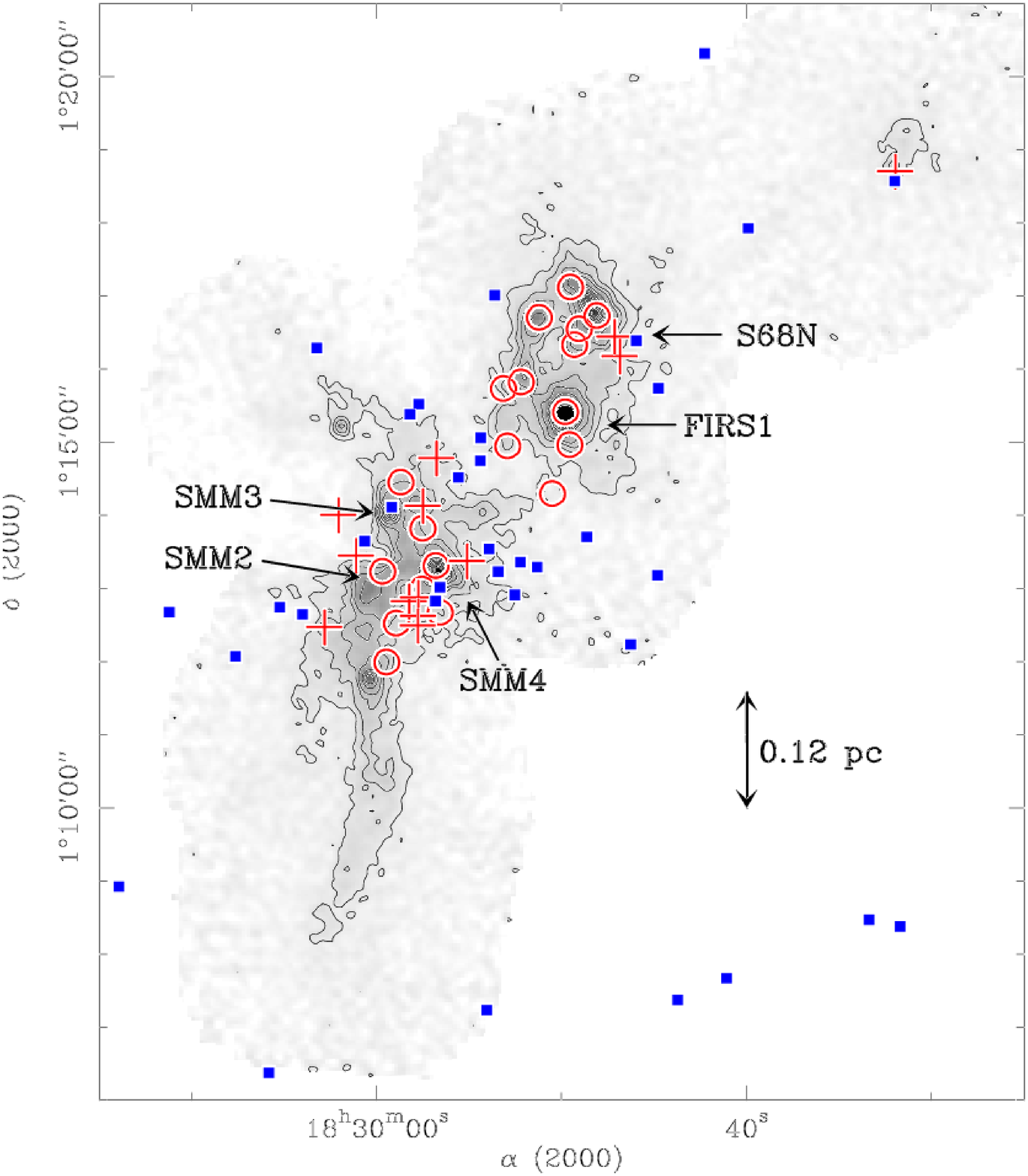}}}
\caption{Dust continuum mosaic (contours and greyscale) of the Serpens main 
core taken at 1.3~mm with the IRAM~30m telescope and the MPIfR 37-channel 
bolometer array (MAMBO). Contour levels: 50. to 200 by 50, 300 to 1000. 
by 100. mJy/11\arcsec -beam; rms noise level: $\sim 17$~mJy/11\arcsec -beam.
The Class\,0 sources are indicated with names and arrows. The location of 
Class\,I sources (red circles), flat-spectrum sources (red crosses), and 
Class\,II sources (blue filled squares) demonstrates the strong clustering of 
protostar candidates along the dense filament.
\label{fig-iram}}
\end{figure*}


\begin{figure}[htb]
\resizebox{\hsize}{!}{\rotatebox{0}{\includegraphics{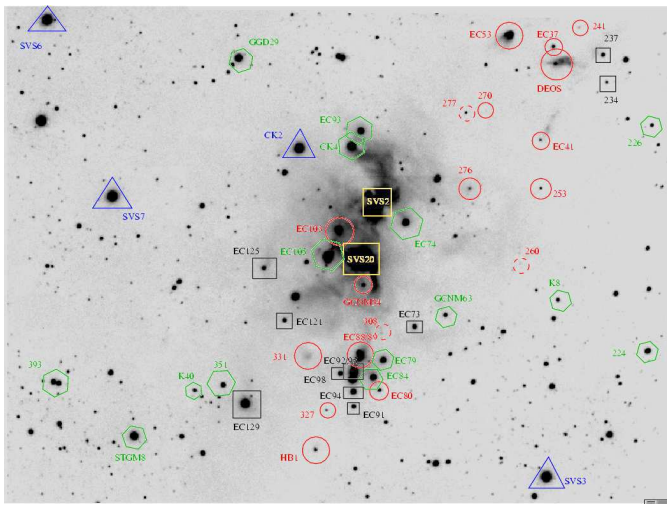}}}
\caption{A $K$ band image of the Serpens Cloud Core (6' $\times$ 8') 
         from \citet{kaa99a} with the Class\,Is (circles), flat-spectrum 
         sources (squares) and Class\,IIs (hexagonals) marked. 
         The ``blue'' ISOCAM sources are marked with triangles. 
         The ridge along SE-NW is well outlined by the Class\,I 
         and flat-spectrum sources.   
         \label{fig-kim}}
\end{figure}

\subsection{Clustering scales}

The concentration of young sources found along the NW-SE ridge, corresponds
well to the pronounced density enhancement seen in the IRAM 1300 $\mu$m map 
in Fig.~\ref{fig-iram}. It is evident that the Class\,I and the flat-spectrum 
sources are spatially more confined to the NW-SE ridge than the Class\,II 
sources, which have a more scattered distribution. In particular, the 
protostars seem to form a set of sub-clusters lined up along the ridge. 
These sub-clusters are in good positional agreement with the kinematically 
separated N$_2$H$^+$ cores A,B,C,D found by \citet{tes00}, especially with 
B,C, and D. The SVS4 sub-cluster lies between cores A and B, however, and is 
also displaced to the west of the SMM4 mm continuum peak.


\begin{figure}[htb]
\resizebox{\hsize}{!}{\rotatebox{270}{\includegraphics{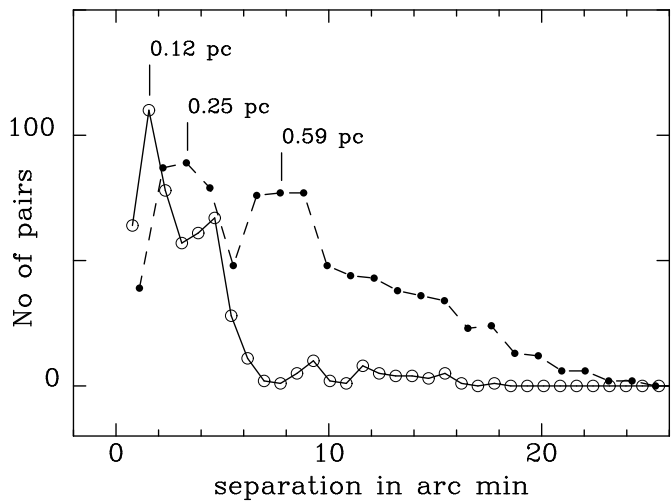}}}
\caption{The distribution of source separations for Class I and flat-spectrum
           objects (circles) and Class\,II sources (dots). The number of 
         separations between sources (per bin) is plotted on the y-axis 
         as a function of separation in arc minutes on the x-axis. The 
         chosen bin size was 6 times the minimum separation, translating 
         to 46.1$\arcsec$ and 66.1$\arcsec$ for the protostar and Class\,II 
         source samples, respectively. 
         \label{fig-clust}}
\end{figure}

The spatial distribution of all the classes of YSOs within the central 
6' $\times$ 8' of the Cloud Core can be seen overlaid on a $K$ band image 
in Fig.~\ref{fig-kim}. To quantify the scale of sub-clustering of
the two populations: protostar candidates (Class\,I and flat-spectrum 
sources) and Class\,II sources, we show for each population the distribution 
of separations between the sources.
Figure~\ref{fig-clust} shows the result as the number of pairs vs. separation.
The bin size was chosen to be 6 times the minimum separation. The difference 
between the two populations is clear. The protostar candidates (circles) have 
a minimum scale of clustering of about 0.12 pc (95$\arcsec$), a secondary
peak at 0.36 pc, and practically no scattered distribution, while the 
Class\,II sources have a minimum clustering scale of about 0.25 pc, a secondary
peak at 0.59 pc and a much more distributed population, in agreement with the 
maps. The clustering scale of 0.12 pc for the protostars agrees well with the 
range of radii of the N$_2$H$^+$ cores: 0.055 to 0.115 pc \citep{olm02}, 
and the second peak at about 0.4 pc reflects well the size scale of the most 
active region in the Serpens Core.

The correspondence between the high density gas and the compactness of the
protostar clusters indicates that these sources are found close to their
birth place. Assuming a typical velocity dispersion in the range from 0.1 
to 1 km\,s$^{-1}$, YSOs younger than $10^5$ yrs can be expected to have 
moved at most $0.1$ pc from their birth place. The typical ages of Class\,I 
sources are estimated to be of this order in other regions such as $\rho$ 
Ophiuchi \citep{gre94} and Taurus \citep{ken90b}. Our results indicate that
the protostar candidates in each cluster must have formed from the same
core at about the same time, which may put constraints on models of cloud
fragmentation and core formation within clumps. Each sub-cluster contains 
between 6 and 12 protostar candidates, and this is just a lower limit because 
of the spatial resolution of ISOCAM. Adopting the Serpens distance of $260$ 
pc yields a {\em protostar} surface density in the sub-clusters in the range 
from 500-1100 pc$^{-2}$. To our knowledge, such a high spatial density of 
protostars has not been found in any other nearby star formation region, 
although we note that the Class\,0 surface density in Orion-OMC3 is 
comparable to that in the Serpens Core \citep{chi97}.

The sub-clustering in the observed 2D projection of the cluster does not 
necessarily correspond to the true spatial distribution of the objects, but 
it seems highly unlikely that these concentrations are due to elongated cloud 
structures seen end-on. Sub-clustering was also found for the Class\,II
sources in $\rho$ Ophiuchi \citep{bon01}, with a similar size scale as
for the Class\,II population in Serpens. It seems likely that the strong
clustering of Serpens protostars will evolve into looser clusters of 
Class\,IIs over a few 10$^6$ yrs. The expansion of the clustering scales 
from 0.12 to 0.25 pc and the assumed ages of these two populations 
suggest that the velocity dispersion of the young stars is of the order of 
only 0.05 km\,s$^{-1}$.

\section{Star formation rate and efficiency}

Assuming that a recent burst of star formation took place $\sim 2 \times$ 
10$^5$ yrs ago, producing the current population of protostars, we can derive 
a rough estimate of the star formation rate (SFR) and the star formation 
efficiency (SFE) in this burst. 
We use our sample of protostar candidates, the flat-spectrum and Class\,I 
sources found clustered along the NW-SE oriented ridge. We are not able to 
estimate their masses, but we hypothesize that they follow the same 
IMF as the Class\,II sample. Comparing the total mass of the Class\,II 
sample and correcting for the number of sources, we estimate a total mass 
of 12.1 $M_{\odot}$ for the protostar population. If these sources were 
formed gradually over $2 \times$ 10$^5$ yrs, the SFR in this microburst 
would be 6.1 $\times$ 10$^{-5}$ M$_{\odot}$/yr, or for typical masses of 
0.17 M$_{\odot}$, one newborn star every $\sim$ 2800 yr. This SFR is  
significantly higher than the one found for the whole YSO population 
in $\rho$ Ophiuchi \citep{bon01}. 

Comparing the protostellar masses with the gaseous mass in these cores 
will give us the star formation efficiency, defined as $SFE = 
M_{\rm star}/(M_{\rm star}+M_{\rm gas})$. According to \citet{olm02} the 
two sub-clumps they name NW and SE, which contain our four sub-clusters of 
protostars, are virialised with masses of 60 $M_{\odot}$ each. For our
protostar candidates this yields a SFE of about 9 \%. 
This is the {\em local} SFE for the sub-clusters along the NW-SE ridge. 
The global SFE calculated for the Class\,IIs and protostars over the entire 
cluster (i.e. 28.8 $M_{\odot}$ of stellar mass) is much lower. In the 
literature we find estimates of the total gas mass from 300 $M_{\odot}$ 
\citep{mcm00} to 1500 $M_{\odot}$ \citep{whi95} for surveyed areas of 5
to 10 arc minutes, which gives only upper limits on the SFE of the order 
of 2-9 \%.

The above estimates are based on the hypothesis that the protostars follow 
the same IMF as the one found for the Class\,II sources. We have not 
attempted to derive the protostar mass function. Future high resolution 
far-IR observations (e.g. ESA's Herschel Space Observatory) would be needed 
to yield bolometric luminosities of these clustered protostars, and with some 
assumptions on the accretion rates one could estimate their masses.


\section{Summary and conclusions}

We have used ISOCAM to survey 0.13 square degrees of the Serpens Cloud 
Core in two broad bands centred at 6.7 and 14.3 $\mu$m. In combination
with our ground based deep $JHK$ imaging of the 48 square arcminute 
central region as well as additional $K$ band imaging of about 30 square
arcminutes to the NW, and a 1.3 mm IRAM map of the central dense filament, 
we have investigated the mid-IR properties of the young stellar population. 
The following results are found:

\begin{itemize}
\item The number of point sources with reliable flux measurements are 
      392 at 6.7 $\mu$m and 139 at 14.3 $\mu$m. Of these, 124 are detected
      in both bands.

\item On the basis of one single colour index, $[14.3/6.7]$, we found that
      53 of the 124 objects possess strong mid-IR excesses. Only 28 of 
      these were previously suggested YSO candidates. The large scale 
      spatial distribution of these mid-IR excess sources is strongly 
      concentrated towards the Cloud Core, where it is elongated along 
      NW-SE.

\item The near-IR $J-H/H-K$ diagram is found to have an efficiency of less
      than 50\% in detecting IR excess sources. This efficieny is comparable 
        to that one found in star formation regions in general from ISOCAM data
        \citep{kaa00} and $\rho$ Ophiuchi in particular \citep{bon01} and 
      should be kept in mind when interpreting $JHK$ based data in terms of 
      disk fractions.

\item The $H-K/K-m_7$ diagram separates well intrinsic IR excess from
      the effects of reddening. From this diagram we were able to increase
      the number of mid-IR excess sources to 70. This means a fractional 
      increase in the investigated region by 25\%.

\item Combination of near-IR and mid-IR photometry for reddened stars 
      without IR excesses enables us to estimate the extinction at 6.7 and 
      14.3 $\mu$m relative to that in the $K$ band in the Serpens direction. 
      We find A$_{7}$ = 0.41 A$_K$ and A$_{14}$ = 0.36 A$_K$.  
      Our results agree with the extinction law measured towards the Galactic 
      centre by \citet{lut96}, as well as with the results from the ISOGAL survey 
      \citep{jia03}.

\item Classification of the Serpens YSOs in terms of the SED indices 
      gives 20 Class\,I sources, 13 flat-spectrum sources, and 43 Class\,II 
      sources. The number of Class\,I sources appears to be 
      exceptionally large along the NW-SE oriented ridge, and the number 
      fraction Class\,I/Class\,II is almost 10 times higher than normal, an 
      indication that this part of the cluster is extremely young and active.

\item The mid-IR luminosities of the Class\,I sources are on the average
      larger than those of the Class\,II sources. Since Class\,II sources
      are expected to have SEDs which peak approximately in the near and 
      mid-IR, in contrast to Class\,Is which radiate most of their 
      luminosity in the far-IR, we conclude that there is a weak indication 
      of luminosity evolution in the Serpens Cloud Core.

\item We have estimated extinction and stellar luminosities for the 43 
      Class\,II sources found in our survey. The Class\,II luminosity
      function is found to be compatible with co-eval formation about 
      2 Myrs ago and an underlying IMF of the three-segment power-law
      type \citep{kro93,sca98}, similar to the mass function found in
      $\rho$ Ophiuchi \citep{bon01}. With this assumption on age, every
       fifth Class\,II is a young brown dwarf.

\item Except for one case, the Class\,I sources are exclusively found in 
      sub-clusters of sizes $\sim$ 0.12 pc distributed along the NW-SE 
      oriented ridge. The sub-clusters also contain several flat-spectrum 
      sources. In total, each core has formed between 6 and 12 protostars 
      (lower limit) within a very short time. 
      The spatial distribution of the Class\,II sources, on the other hand, 
      is in general much more dispersed. 

\item On the assumption that the protostar candidates follow roughly the
      same IMF as the Class\,IIs, we derive a SFR of 6.1 $\times$ 10$^{-5}$ 
      M$_{\odot}$/yr and a local SFE of 9\% in the recent microburst of
      star formation forming the dense sub-clusters of protostars.

\end{itemize}

The results presented in this study show evidence that the sub-clusters in 
the central part of the Serpens Cloud Core were formed by a recent microburst 
of star formation. The extreme youth of this burst, deduced from the compact
clusters of protostar candidates, is supported by independent investigations, 
such as the rich collection of Class\,0 sources found by \citet{cas93} and 
\citet{hur96} in the same regions.  
In addition to the clustered protostar population, we also find a more 
distributed population of Class\,II sources, for which we have deduced an age 
more or less typical of Class\,IIs found in other regions (2 Myr). In addition 
to these YSO generations, there is probably also a population of Class\,III 
sources, which is undistinguishable from the field star population in our study, 
but which should be looked for with proper search tools, such as e.g. X-ray 
mapping \citep{gro00}. 

\citet{cas93} suggested that their submm sources without near-IR counterparts 
represented a second phase of active star formation in Serpens. Here we show 
the co-existence of sources in the various evolutionary stages from Class\,II 
and flat-spectrum to Class\,I and Class\,0 sources. While we find an age 
estimate of 2 Myr for the Class\,II population, it is highly unlikely that 
the clustered Class\,I and flat-spectrum sources are older than a few 10$^5$ 
yrs. Our results therefore support the conclusions of \citet{cas93} that star 
formation has proceeded in several phases in Serpens.

\begin{acknowledgements}
The authors wish to thank Carlos Eiroa for stimulating discussions and helpful 
comments. We also thank an anonymous referee for comments that helped us 
improve the paper.
The ISOCAM data presented in this paper were reduced using "CIA", a joint 
development by the ESA Astrophysics Division and the ISOCAM Consortium led 
by the ISOCAM PI, C. Cesarsky, Direction des Sciences de la Mati\`{e}re, 
C.E.A., France. The near-IR data presented in this paper were obtained with 
the ARcetri Near InfraRed CAmera ({\sc Arnica}) at the Nordic Optical 
Telescope in 1996, and Carlo Baffa, Mauro Sozzi, Ruggero Stanga and Leonardo 
Testi from the {\sc Arnica} team are acknowledged for the instrument support. 
The Nordic Optical Telescope is operated on the island of La Palma 
jointly by Denmark, Finland, Iceland, Norway, and Sweden, in the Spanish 
Observatorio del Roque de los Muchachos of the Instituto de Astrofisica de 
Canarias. This publication made use of the SIMBAD database, operated at CDS, 
Strasbourg, France, and data products from the Two Micron All Sky Survey, 
which is a joint project of the University of Massachusetts and the Infrared 
Processing and Analysis Center/California Institute of Technology, funded 
by the National Aeronautics and Space Administration and the National Science 
Foundation. Financial support from the Swedish National Space Board is 
acknowledged.

\end{acknowledgements}
%

\bibliographystyle{aa}


\appendix

\section{Notes on some individual sources}
\label{ind}

\subsection{IRAS 18269+0116 (ISO-159, 160)}

At the location of IRAS 18269+0116 we find ISO-159 and ISO-160, of 
which the latter is not detected at 14.3 $\mu$m. A $K$-band image shows that 
ISO-159 has a bow-shaped nebulosity 7$\arcsec$ to WNW (at a level of about 
3-4 $\sigma$, see Fig.~\ref{fig-nw}). The SED index classifies ISO-159 as a 
flat-spectrum source, while ISO-160 is either an extremely extinguished 
background source (analogous to CK2) or most likely a Class\,II object 
(cf Fig.~\ref{fig-k2k3}).

\subsection{IRAS 18271+0102 (ISO-221)}

Identified as a YSO by \citet{cla91} on the basis of its steeply rising SED 
towards longer wavelengths. With ISOCAM we see two very bright point sources
(221 and 238), of which only ISO-221 has IR-excess.

\subsection{EC21 (ISO-231)}

This source was suggested as a YSO on the basis of its location in the $K$ vs
$H-K$ diagram by \citet{eir92}. Deeper near-IR imaging detected the source 
also in the $J$ band, but found no near-IR excess \citep{kaa99a}. It is
located in the very dense and active S68N region, but whether it has intrinsic
IR-excess at 7 $\mu$m or is only reddened (in that case, by 3 magnitudes in the
$K$ band, which translates to $A_V \sim 30$ magnitudes), is not clear. 

\subsection{EC23 (ISO-232)}

Proposed as a YSO on the basis of its $H-K$ colour and 3.08 $\mu$m ice 
feature \citep{eir92}, but not found to have near-IR excess in a deeper 
JHK study \citep{kaa99a}, nor mid-IR excess in this study.

\subsection{IRAS 18272+0114 (DEOS, EC37, EC38, EC53, 234, 237, 241)}

IRAS 18272+0114 is resolved into a cluster of protostar candidates: the 
Deeply Embedded Outburst Source \citep{hod96}, EC37, EC38 and EC53 
\citep{eir92}, and three new ISOCAM sources (234, 237, 241). ISO-241 is
within the positional uncertainties associated with S68N, a radio source 
detected by \citet{mcm94} and found to satisfy the Class\,0 criteria.
We use the position at its 450 $\mu$m emission peak \citep{wol98}. 
ISO-241 is a Class\,I, however, from the shape of its SED from 2 to 14 
$\mu$m. \citet{hod99} denotes the $K$-band source SMM9-IR, since it is 
located close to the SMM9 positions quoted in the literature 
\citep{whi95,wol98,tes98,dav99}, and shows that it has a very faint $K$ 
companion about 2$\arcsec$ to the NE.

\subsection{EC41 (ISO-258a) and FIRS1 (ISO-258b)}

IRAS 18273+0113, also called FIRS1 and SMM1 \citep{rod89,eir89,cas93}, is 
associated with 5 mid-IR excess sources (EC41, 253, 270, 276, 277) plus one 
14.3 $\mu$m source (258b). 

\begin{figure}[htb]
\resizebox{\hsize}{!}{\rotatebox{0}{\includegraphics{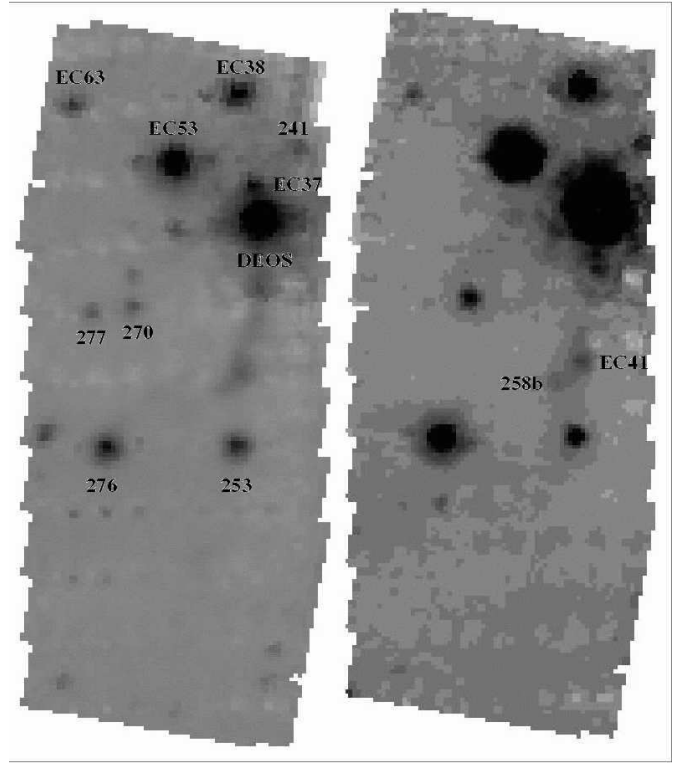}}}
\caption{The deep map D2 at 6.7 $\mu$m (left) and 14.3 $\mu$m (right). North 
up and East left. \label{fig-D2}}
\end{figure}

ISO-Ser-258a is in agreement with the near-IR source positions of EC41 
\citep{eir92} and GCNM23 \citep{gio98}, and these are most likely the same 
source, a Class\,I YSO according to the $\alpha_{\rm IR}^{2-14}$ index. 
The ISOCAM source appeared slightly extended towards the SE both at 6.7 
and 14.3 $\mu$m in most of our data. \citet{lar00} found evidence for two 
separate sources, however, tracing pixel by pixel the 5 to 15 $\mu$m SED 
from CVF observations. This prompted us to take a more careful look, and 
we identified two clearly separated sources at 14.3 $\mu$m (rather than one 
extended source) in one of the deeper (and higher resolution) survey maps, 
D2 (cf. Fig.~\ref{fig-D2}). The new source (called 258b), is located about 
13$\arcsec$ to the SE of EC41. Since the radio position at RA(2000) 
= 18 29 49.73, DEC(2000) = 01 15 20.8  \citep{cur93} is 8.6$\arcsec$ 
to the west of 258b we find it unlikely that this source is the mid-IR 
counterpart of the far-IR source FIRS1, but it could be scattered light we 
see in the mid-IR. It is not entirely clear if this source is just a knot of 
the extended emission which appears to begin somewhat south of DEOS and is 
seen in both filters in Fig.~\ref{fig-D2}. The northernmost part of this 
extended emission is, however, disturbed by a memory feature from the strong 
DEOS source in the raster scanning.


\subsection{IRAS 18273+0059 (ISO-259)}

A prominent extended emission seen in the ISOCAM images corresponds in position 
to IRAS 18273+0059 \citep{cla91}. With ISOCAM, however, we see no apparent 
bright point source embedded in the nebulosity, but a moderately faint object 
is detected at 6.7 $\mu$m only. On optical plates the SE part of this extended 
emission is recognised as an optical reflection nebula. 

\subsection{IRAS 18274+0112}

IRAS 18274+0112 is also composed of multiple mid-IR excess sources centred 
on SVS20 (cf. Fig~\ref{fig-kim}). Higher spatial 25 $\mu$m resolution was 
obtained by \citet{hur96}, and their sources PS1 and PS2 (EC129) are among 
the bright mid-IR excess sources resolved with ISOCAM.

\subsection{EC59 (ISO-272)}

Included as a YSO candidate solely on the basis of its location in the
$K$ vs $H-K$ diagram \citep{eir92}, and no near-IR excess found in a deeper
JHK study \citep{kaa99a}. Since we have no indication of mid-IR excess its 
status remains inconclusive.

\subsection{EC69 {ISO-289}}

This source (also called CK10) was suggested as a YSO on the basis of its 
location in the $K$ vs $H-K$ diagram \citep{eir92}, and while near-IR excess
was found by \citet{sog97}, no near-IR excess was found in a deeper JHK study 
\citep{kaa99a}. The J-band magnitude has varied though: 15.7 in 1989, 14.96 
in May 1992, 15.6 in Nov 1992, and 16.15 in Aug 1996 
\citep{eir92,sog97,gio98,kaa99a}. We conclude that it is a YSO, and with an
$\alpha_{\rm IR}^{2-7} = -1.42$ it belongs in the Class\,II group.

\subsection{EC95 \& EC92 (ISO-317)}

ISO-317 is located at the position of EC92 according to the coordinates 
given by \citet{eir92}, but the positional uncertainty is roughly $\pm 
3\arcsec$. Also, the spatial resolution of ISOCAM is not sufficient to 
resolve the two neighbours EC92 and EC95, and it is likely that both are 
included in the ISOCAM fluxes. The mid-IR SED suggests a flat-spectrum 
source. Slightly closer to the position of EC95, but also with some
positional uncertainty, \citet{pre98} found an extremely strong X-ray 
source (Ser-X3).

\subsection{ISO-331}

This bright mid-IR source is detectable in $K$ as a nebulous spot only. 
Probably the $K$ flux is dominated by scattered light, and it is therefore 
not entirely correct to place ISO-331 in colour diagrams together with 
continuum sources. Nevertheless, the $K$ measurement gives a lower limit to 
the colour index of the object, which is classified as a Class\,I source. 
Its mid-IR position is about 25\arcsec to the NNW of SMM2 \citep{cas93}, 
and coincides roughly with the VLA 3.6 cm continuum source \#11 found by 
\citet{bon96}.

\subsection{CK2 (ISO-337)}

According to ISOCAM data the source CK2, which is extremely red in the 
near-IR, has no mid-IR excess and can be interpreted as a background source, 
consistent with the suggestions of \citet{chu86}, \citet{chi94}, \citet{cas96}
and \citet{sog97}. The extinction towards CK2 is high, A$_V >$ 50 mag.

\end{document}